%% file: main.tex
\documentclass{ifacconf}

\usepackage[usenames,dvipsnames]{pstricks}
\usepackage{epsfig}
\usepackage{mathtools}
\usepackage{palatino}
\usepackage{tikz}
\usetikzlibrary{shapes,shapes.geometric,arrows,fit,calc,positioning,automata,}
\RequirePackage{ifthen}
\usepackage{latexsym}
\RequirePackage{amsmath}
\RequirePackage{amssymb}
\RequirePackage{xspace}
\RequirePackage{graphics}
\usepackage{float}
\usepackage{xcolor}
\usepackage{keyval}
\usepackage{xspace}
\usepackage{mathrsfs}
\usepackage{graphicx}
\usepackage{paralist}
\usepackage{array}
\newcolumntype{P}[1]{>{\centering\arraybackslash}p{#1}}
\usepackage{makecell}
\usepackage[algo2e,ruled,commentsnumbered,linesnumbered,noend]{algorithm2e} 
\usepackage{amsmath,amssymb,url,listings,mathrsfs}
\usepackage{multirow}
\usepackage[americancurrent,siunitx]{circuitikz}
\usepackage{balance}
\usetikzlibrary{positioning}
\usepackage{natbib}        % required for bibliography
\usepackage{siunitx}
\usepackage{balance}

\tikzstyle{label}=[draw, circle, outer sep=6pt,
  inner sep=1pt, align=center]
\newcommand{\myTrans}[1]{\tikz \node[label, font=\scriptsize, outer sep=0pt] {$#1$};}

\newcommand{\vdd}{V_{DD}}

\newcommand{\myTime}[1]{\SI{#1}{\second}}
\newcommand{\myTimeS}[1]{\SI{#1}{\second}}

\newcommand{\tool}[1]{\textit{#1}}
\newcommand{\reals}{{\mathbb R}}                    %reals
\newcommand{\Reach}{{\mathtt{Reach}}}
\newcommand{\vertiii}[1]{{\left\vert\kern-0.25ex\left\vert\kern-0.25ex\left\vert #1 
    \right\vert\kern-0.25ex\right\vert\kern-0.25ex\right\vert}}

\newcommand{\interval}{}

\newcommand{\chull}{{\mathtt{hull}}}

\newcommand{\C}{\mathcal{C}}

\newcommand{\R}{\mathcal{R}}

\newcommand{\TB}{\mathsf{STB}}
\newcommand{\unsafe}{\mathtt{unsafe}}

\newcommand{\nnreals}{\mathbb R_{\geq 0}}

\newcommand{\mur}[1]{\multirow{2}{*}{#1}}

\newcommand{\U}{\mathcal{U}}

\newcommand{\dia}{\mathit{dia}}
\newcommand{\InvHy}{\sf{InvHy}}
\newcommand{\InvUni}{\sf{InvUni}}
\newcommand{\InvLoop}{\sf{InvLoop}}

\newcommand{\NOR}{\sf{NOR}}
\newcommand{\OR}{\sf{OR}}
\newcommand{\sig}{\sf{Sig}}
\newcommand{\ramp}{\sf{Ramp}}
\begin{document}
\begin{frontmatter}
\title{Verifying nonlinear analog and mixed-signal circuits with inputs}

\author[First]{Chuchu Fan}
\author[First]{Yu Meng}
\author[Second]{J{\"u}rgen Maier}
\author[Second]{Ezio Bartocci}
\author[First]{Sayan Mitra}
\author[Second]{Ulrich Schmid}

\address[First]{University of Illinois at Urbana-Champaign, (emails: \{cfan10, yumeng5, mitras\}@illinois.edu )}
\address[Second]{Technische Universit\"at Wien, (emails: \{jmaier, s\}@ecs.tuwien.ac.at, ezio.bartocci@tuwien.ac.at)}
% \IEEEauthorblockA{\IEEEauthorrefmark{1}
%{\{cfan10, yumeng5, mitras\}@illinois.edu}\\
%University of Illinois at Urbana-Champaign
%} \IEEEauthorblockA{\IEEEauthorrefmark{2}
%{\{jmaier, s\}@ecs.tuwien.ac.at, ezio.bartocci@tuwien.ac.at}\\
%Technische Universit\"at Wien
%}

\begin{abstract}
%There has been progress in verification of nonlinear and hybrid systems in the recent years using algorithms that combine simulation data with model-based sensitivity analysis. These approaches only handle closed models, that is, models without inputs. The na\"{i}ve introduction of models of input signals breaks these approaches, as typical inputs (fast sigmoids, discontinuous functions) for analog and mixed-signal circuits make the system highly sensitive and the number of needed simulations grow rapidly. In this paper, 
We present a new technique for verifying nonlinear and hybrid models with inputs.  
We observe that once an input signal is fixed, the sensitivity analysis of the model can be computed much more precisely. Based on this result, we propose a new simulation-driven  verification algorithm and apply it to a suite of nonlinear and hybrid models of CMOS digital circuits under different input signals. The models are low-dimensional but with highly nonlinear ODEs, with nearly hundreds of logarithmic and exponential terms. Some of our experiments analyze the metastability of bistable circuits with very sensitive ODEs and rigorously establish the  connection between metastability recovery time and sensitivity.
%Our results not only demonstrate the feasibility of our approach but also provide interesting insights like the close connection between metastability recovery time and sensitivity.
\end{abstract}
\end{frontmatter}
%\maketitle

\input{intro}
%
\input{theory}
\input{models}

\input{res}

%\input{other}

\input{conclusion}
\balance
\section*{Acknowledgments}

We would like to thank the developers of CORA (especially Matthias Althoff),
dReach and Flow* for their support while porting our models.

\bibliography{ref}

%\newpage
%\input{appendix}

\end{document}

%% file: intro.tex
%!TEX root = main.tex
\section{Introduction}
\label{sec:intro}

Analog and mixed-signal circuits have provided  a well-spring 
of hard problem instances for formal verification of hybrid systems (HS). 
%The former has been a well-spring of hard problem instances and 
%the latter has pushed the envelope of the rigorous analysis of complex 
%properties of these circuits. 
%Examples range from various analog and
%mixed-signal verification problems~\cite{DDM04:FMCAD} to metastability 
%analysis of digital circuits like arbiters~\cite{coho2008}.
Tools like HyTech~\citep{henzho:Hytech}, PHAVer~\citep{Fre08},
SpaceEx~\citep{helicopter}, Checkmate~\citep{GKR04:ICCAD}, 
$d/dt$~\citep{DDM04:FMCAD}, and Coho~\citep{coho2008} have targeted and successfully
verified  linear dynamical and hybrid models for tunnel-diode oscillators~\citep{Lata2010}, 
 $\Delta\Sigma$ modulators~\citep{GKR04:ICCAD,DDM04:FMCAD}, 
filtered oscillators~\citep{helicopter}, and digital arbiters~\citep{coho2008}.
%important first generation verification algorithms
%were benchmarked using linear and linear hybrid models of circuits: 
%They simulate dense bundles of trajectories or reachable states enabling 
%exploration of all the possible behaviors of the circuit~\cite{GreenstreetM99}. 
%
%However, realistic circuit models are (highly) non-linear, which makes linear 
%models questionable in general, and causes the linear approximation-based 
%methods to be too conservative.  
Only recently, verification tools such as Flow*~\citep{chen2013flow},
NLTOOLBOX~\citep{DangGM09}, iSAT~\citep{FranzleHTRS07},
dReach~\citep{kong2015dreach}, C2E2~\citep{CAVtool} and CORA~\citep{Althoff2016a}, 
have demonstrated the feasibility of verifying nonlinear dynamic and hybrid models. 
These tools are still limited in terms of the complexity 
of the models and the type of external inputs they can handle, and they
require quite often manual tuning of algorithmic parameters.
The verification challenge for nonlinear circuits is further exacerbated by the fact 
that  these problems often require state exploration in regions, 
where the model is very sensitive. For example, bi-stable circuits like a
storage element or a flip-flop can be driven into a metastable state where the
circuit may output signals in the forbidden region between logical $0$ and
logical $1$ or experience very high-frequency oscillations for an  
arbitrary time, before resolving to a proper state~\citep{Mar81}.  

%The nature 
%of this phenomenon suggests that the model is very sensitive in the 
%metastable region, which is indeed confirmed by our approach. 

In this paper, we present a novel technique for verifying  
nonlinear dynamic and hybrid models with externally controlled input functions. 
The approach builds up on  previous work that combines numerical simulation with 
model-based sensitivity analysis for bounded invariant verification~\citep{CAVtool}. 
%Consider a nonlinear ODE $\dot{x} = f(x)$, with a set of initial states 
%$\Theta \subseteq \reals^n$, a time bound $T$, and a set of unsafe states 
%$\unsafe$. A solution of the system is a function $\xi: \reals^n \times \nnreals \rightarrow \reals^n$, 
%such that for any initial state $x_0 \in \reals^n$ and time $t$, $\xi(x_0,t)$ 
%satisfies the ODE. 
For (bounded) invariant verification, we need to check 
whether the set of reachable states (up to a given time $T$) intersects with the unsafe set. 
In general, computing the exact bounded reachable set for nonlinear dynamic systems is hard. 
The simulation-driven  approach circumvents this by over-approximating the reachable states  using numerical 
simulations from a finite number of initial states  and bloating 
these simulations by a factor determined by the sensitivity of the 
solutions to initial states.
%the higher the sensitivity of the system 
%to initial states, the larger is the number of simulations required to compute over-approximations with certain error bounds. 
%
%	If this over-approximation  proves safety or produces a counter-example, then the algorithm decides, otherwise, it draws more samples of initial states and repeats the earlier steps to compute a more precise over-approximation.
Previous work establishes that the resulting algorithms are  sound,  complete for 
robust invariant verification~\citep{duggirala2013verification}, and 
extensible to nonlinear hybrid models~\citep{CAVtool}.
The key ingredient for the effectiveness of this approach is the precise
symbolic approximation of sensitivity as formalized by so called
\emph{discrepancy functions}. If the discrepancy function is excessively
conservative, then in practice, the verification algorithm may trigger many refinements, and
never reach a decision.

%This approach is implemented in the tool C2E2~\cite{CAVtool} and has been used to successfully verify the safety of  power-train control systems~\cite{duggirala2015meeting}, parallel landing protocols, and several other complex nonlinear models.

One major shortcoming of existing approaches is their inability to handle
sensitivity analysis of models with external inputs.
For a typical digital circuit like a simple inverter, the output trajectory
$V_{out}$ depends strongly on the input trajectory $V_{in}$. The na\"{i}ve
approach for handling external inputs,
namely, making the system closed by considering input signals as 
additional state variables, does not work as the resulting discrepancy 
functions become  too conservative to be effective, i.e., the overapproximation
  error of the discrepancy function becomes too large (see the
discussion in Section~\ref{sec:verificationproblem}).

%over-approximations 
%of the reachable sets in the case of highly sensitive systems,
%which often fill up large portion of the state space. 

%We address the sensitivity issue by introducing a new technique to compute discrepancy which separates input from state variables: 
In this paper we therefore propose a new method for computing discrepancy functions for
open models.  We show that for a given input signal and a numerical solution of
the model, it is possible to compute precise upper-bounds on solutions from
neighboring initial states, experiencing the same input
(Theorem~\ref{lemma:final}). Using this new method for discrepancy computation,
we have generalized the verification algorithm to handle nonlinear hybrid models
with inputs. This approach was later integrated in a new version of the tool
C2E2 described in detail in~\cite{CAVtool}. 
% for verifying 
%circuits with fixed inputs  through reach set over-approximations,  preserving soundness and 
%completeness of the original algorithm.
%verification algorithm for nonlinear hybrid models. 
%Owing to this extension, we can stimulate and verify sensitive models with arbitrary input signals. 
We show the feasibility of our approach by evaluating several Complementary
Metal-Oxide Semiconductor (CMOS) digital 
circuit models, including an inverter, a {\NOR}-gate and an {\OR}-gate\footnote{Files can be found at \url{https://publish.illinois.edu/c2e2-tool/gate/}.}\label{modellink}.
%using both ramp and sigmoid inputs, 
%The ramp input has constant derivatives 
%for the up and down slope of the pulse, while the sigmoid input smooths 
%the non-differentiable states by using polynomial differential equations to approximate the sigmoid function. 
Furthermore, two very sensitive models of storage elements, one
consisting of two inverters in a feedback-loop and the other of an {\OR}-gate with its output fed 
back to one of its inputs, are investigated. The latter allows to memorize a rising transition on its 
second input, and is an ideal target for demonstrating the capability of C2E2 
to even predict metastable behavior correctly. Our results demonstrate
that the new C2E2 can indeed be used to verify reachability problems for open
circuits with unprecedented complexity (over one hundred logarithmic and
exponential terms in differential equations).
Experiments confirm what is known about metastable behavior, and provide detailed
insights into the close connection between sensitivity and metastability recovery (time).
Comparisons to leading verification tools, i.e., Flow*, CORA and dReach show
that C2E2 outperforms these approaches not only in speed but also in the amount of
provided features and the maximal ODE complexity that still can be handled. We
thus conclude that our technique is a very powerful, viable approach for simulating 
dynamic open models.

%Based on these observations, we are convinced that the tool will open up a promising
%new avenue for studying such complex, rare and highly transient events in digital
%circuits.

%The rest of the paper is organized as follows: In Section~\ref{sec:thin}, we introduce the model and the verification problem. In Section~\ref{sec:sensitivity}, we discuss a novel approach to compute the discrepancy function for systems with fixed inputs and in Section~\ref{sec:verification_algorithm}, we give a brief review of the simulation-driven verification algorithm using discrepancy. In Section~\ref{sec:models}, we discuss modeling of CMOS circuits, and in Section~\ref{sec:res} we present verification results.
%Conclusions and directions of future research are provided in Section~\ref{sec:conclusion}.

%\sayan{This also belongs to a short ``related work'' para somewhere in the intro.}

%\sayan{I suggest dropping these comparisons for now; unless there is a more logical place to work them in.}
%{Input-to-state (IS) discrepancy functions} as proposed in~\cite{HuangM:HSCC2014}  
%take into consideration the sensitivity w.r.t. {\em different\/} input signals, 
%and the method for computing IS-discrepancy presented in~\cite{chuchuATVA} 
%introduces a multiplicative error factor of $e^{\frac{1}{2}t}$. This makes the 
%over-approximation too conservative to be useful.

%% file: theory.tex
%!TEX root = main.tex
\section{Simulation-Driven Verification}
\label{sec:thin}
\paragraph*{Notations} 
%$\reals$ 
%denotes the set of real numbers. 
For a real vector $x \in \reals^{n}$, $\|x\|$ denotes its $l^2$ norm. For a set ${S}\subset \reals^n$, the diameter  $\dia({S})$ is the supremum of the distance between any pair of points in ${S}$.
For $\delta \geq 0$, $B_\delta(x)$ is the closed $\delta$-ball around $x$.
%, that is, the set $\{x' \in \reals^n  \ | \ ||x' - x || \leq \delta \}$. 
Closed $\delta$-balls around sets are defined %in the natural way.
as $B_{\delta}(S) = \cup_{x \in S}B_\delta(x)$. 
$S \oplus S'$ is the Minkowski sum of  sets $S$ and $S'$.
%, where $B_{\delta}(x)$ denotes the set $\{y|\|x-y\|\leq \delta\}$. 
For a real matrix $A \in \reals^{n \times n}$, $(A)_{ij}$ is the entry on the $i^{th}$ row and the $j^{th}$ column; $eig(A)$ is the largest eigenvalue of $A$.
%\sayan{strange notaion. usually $a_{ij}$ are the entries of $A$. $[A]$ makes it confusing.}
%
For a pair of matrices $\underline{A},\bar{A}$ with $(\underline{A})_{ij} \leq (\bar{A})_{ij}$ for all $1\leq i,j \leq n$, we define the {\em interval matrix} as the set of matrices: 
\begin{eqnarray*}
\interval [\underline{A},\bar{A}]  \triangleq  \{A \in \reals^{n \times n} | (\underline{A})_{ij} \leq (A)_{ij} \leq (\bar{A})_{ij}, 1\leq i,j \leq n\}.
\end{eqnarray*} 
%\sayan{Do we need $interval([\underline{A},\bar{A}])$ or just $interval(\underline{A},\bar{A})$ or $[\underline{A},\bar{A}]$? Why different font for interval? }

\begin{figure}[t!]
	\label{fig:simu}
	\centering
	\begin{minipage}{0.15\textwidth}
	\includegraphics[width=\textwidth]{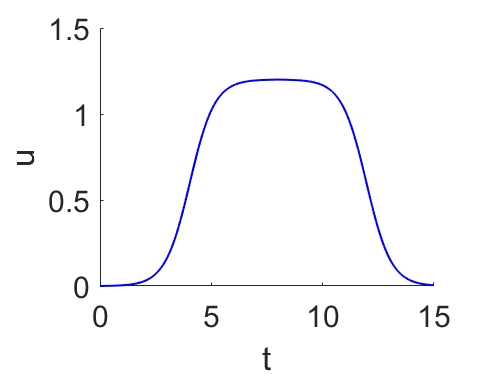}
	\end{minipage}
	\begin{minipage}{0.15\textwidth}
	\includegraphics[width=\textwidth]{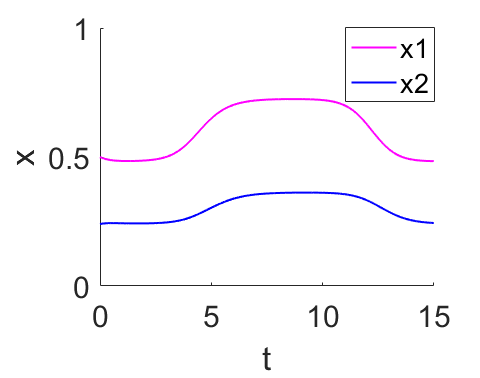}
	\end{minipage}
	\begin{minipage}{0.15\textwidth}
	\includegraphics[width=\textwidth]{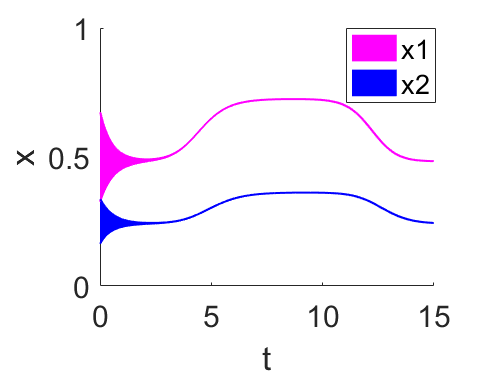}
	\end{minipage}
	\caption{\small Input signal $u(t)$ ({\em left}), corresponding trajectory of $x_1,x_2$ ({\em center}), and reach sets from $B_{0.1}([0.5,0.24]^T)$}
	\label{fig:example}
\end{figure}
%\sayan{Can we also plot $x2$ in the same figure?}

\subsection{Dynamic systems with inputs}
\label{sec:dynamicalsys}
An $n$-dimensional {\em dynamic system\/} with $m$-dimensional {\em input} is described by an
ordinary differential equation:
\begin{eqnarray}
 \dot{x}(t) = f(x(t), u(t)),
 \label{eq:system}
\end{eqnarray}
where $f:\reals^n \times \reals^m \rightarrow \reals^n$ is a continuously differentiable function,
and a compact set $\Theta \subseteq \reals^n$ of {\em initial states}.
The input is an integrable function $u: [0,\infty) \rightarrow \U$, where $\U \subset \reals^{m}$ is a compact set.
% \ulrich{Later on, you require it to be compact - shouldn't we replace bounded by compact also here?}
Given an input  $u$, the  {\em solution\/} or the {\em trajectory\/} of the system is a function $\xi_u :\reals^n \times \reals^m \times \reals_{\geq 0} \rightarrow \reals^n$, such that for any initial state $x_0 \in \Theta$ and at any time $t \in\reals_{\geq 0}$,  $\xi_u(x_0, t)$ satisfies~\eqref{eq:system}. 
%The {\em initial set} of the system \eqref{eq:system} is defined as the set of initial states, and we will denote it as $\Theta$ throughout the paper. 
%Existence and uniqueness of the solution follow from the continues differentiability of $f$.
%
A state $x \in \reals^n$ is {\em reachable} if there exists $x_0 \in \Theta$ and a time $t \geq 0$ such that $\xi_u(x_0,t) = x$.  The set of all reachable states over an interval of time $[0,t_1]$  with input $u$ 
is denoted by $\Reach_{u} (\Theta, [0,t_1])$; $\Reach_{u} (\Theta, [t_1,t_1])$ is written as $\Reach_{u} (\Theta, t_1)$ in brief.
%\sayan{Why $u$ is a subscript and not an argument like the other parameters?}
\begin{exmp}
\label{example:running}
Consider a cardiac oscillator  described by the time-invariant ODEs $\dot x_1= - x_1(x_1^2 + 0.9x_1 + 0.9) + 2 x_2u +1 ; \dot x_2 = x_1 - 2x_2$. For a smoothed sigmoidal input $u$,  
%is described by differential equation: $\dot u = u(1.8-1.5u)+0.0015$ for an the up ramp  and $\dot u = -u(1.8-1.5u)-0.0015$ for the down ramp (see Figure~\ref{fig:example} ({\em left})). 
the corresponding trajectories and  (over-approximations of) reach sets projected on $x_1(t),x_2(t)$ are shown in Figure~\ref{fig:example}.
\end{exmp}

\begin{figure}[t!]
	\centering
	\includegraphics[width=0.20\textwidth,trim={0 0 0 1.5cm},clip]{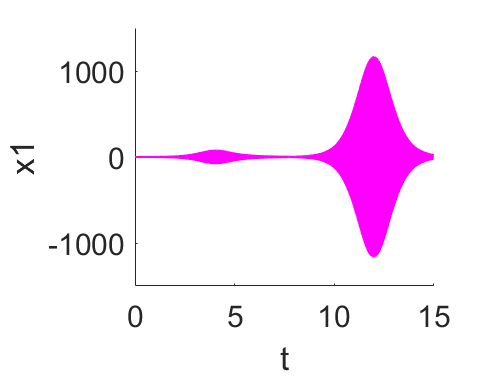}
	\caption{\small Over-approximation of $x_1(t)$ without separate input
          from state variables. Note the different scale compared to
Figure~\ref{eq:system}.}
%\juergen{Add another signal here? e.g. input signal}}  US: I do not think so.
	\label{fig:bloat_up}
\end{figure}

%\sayan{Define reachable set here. State the safety verification problem here. This is also a good place to introduce the running example. I have just moved a para from later section here.}
\subsection{Safety verification problem}
\label{sec:verificationproblem}
Given an $n$-dimensional dynamic system, an input signal $u(t)$, a compact initial set $\Theta \in \reals^n$, an unsafe set $\unsafe \subseteq \reals^n$, and a time bound $T>0$, the safety verification problem is to check
whether $\Reach_{u}(\Theta,[0,T])  \cap \unsafe = \emptyset$.
  
Safety verification of nonlinear ODEs and hybrid models is difficult even in the absence of inputs. 
%(see Section~\ref{sec:related} for a brief overview of related work). 
For closed models (without inputs), recently developed {\em simulation-driven verification algorithms} decide the safety verification question rigorously by combining numerical simulations with sensitivity analysis of the trajectories with respect to their initial states ~\citep{donze2010breach,duggirala2013verification,CAVtool}.
These approaches are most effective when the sensitivity of the solutions to initial states can be precisely approximated. 

These existing approaches do not support models with inputs. The seemingly natural idea of explicitly modeling the input $u$ as a state variable, i.e., its controlling ODE, and then verifying the  resulting closed model does not work. This is because  
inputs often model unstable signals---like the pulse $u$ in Example~1---and in such cases the trajectories of the resulting closed system will turn out to be extremely sensitive with respect to the initial states and render simulation-driven verification ineffective.
%as the resulting closed system can be unstable as the input variable often models unstable behaviors like stiff changes in the pulse.
%\sayan{ Explain this better. The input signal itself might be unstable and cause the entire system to be unstable.} 
In Example~\ref{example:running}, if we treat $u$ as a state variable, the over-approximation reach set of $x_1(t)$  using C2E2 is shown in Figure~\ref{fig:bloat_up}. The (prohibitive) blow-up in the over-approximation is due to the unstable input $\dot u = u(1.8-1.5u)+0.0015$ that models the rising transition of the smoothed pulse.
%\sayan{Say why this happens beyond an empirial observation.}

%\sayan{Talk about determinism in input signal.}
%This motivates us to separate input from state variables and to come up with methods to measure the sensitivity of the system with fixed inputs as described in the next section. 
%\chuchu{Moreover, we set the input variable to be deterministic to avoid an addition error discussed later.}

\section{Sensitivity Analysis for Open Systems}
\label{sec:sensitivity}
%\sayan{This section is about introducing basic pieces needed for analysis, this should come after the problem statement.}
We formalize  sensitivity using  {\em discrepancy functions} as introduced in~\cite{duggirala2013verification}.
%
%\subsection{Discrepancy function for fixed input}
%\label{sec:discrepancy}
%
%
%\sayan{Why not define discrepancy right here. Thenotion of discrepancy we want for this paper, i.e. Then we can relate it to previous notions of discrepancy. Presumably this is more general than the prvious notions?}
%
%Sensitivity of the system is  analysis can be formalized as {\em discrepancy functions}~\cite{}. The techniques proposed in~\cite{ATVA,EMSOFT} to compute discrepancy functions are for dynamical systems without input signals. 
%
Given an input signal $u(t)$ for~\eqref{eq:system}, a discrepancy function bounds the distance between two neighboring trajectories, as a function of the  distance between their initial states and the time. 
%That is, given any two trajectories $\xi_u(x,t)$ and $\xi_u(x',t)$ of the system~\eqref{eq:system} starting from states $x$ and $x'$, respectively, with input $u(t)$, the discrepancy function $\beta_u$ is a function of the distance between $x$ and $x'$, and time $t$. The distance between $\xi_u(x,t)$ and $\xi_u(x',t)$ is upper-bounded by the discrepancy function at $t$:
\begin{defn}
\label{def:pdf}
Given an input signal $u(t)$, a continuous function 
%parameterized by  $u$,
%sayan:  $u$ is not a parameter of \beta formally; $\beta$ depends on $u$ and that is amply clear 
% from the notation $\beat_u$
 $\beta_u:\nnreals \times \nnreals \rightarrow \nnreals$ is a {discrepancy function} of~\eqref{eq:system} if
\begin{enumerate}[(1)]
\item for any pair of states $x, x' \in \reals^n$, and any time $t \geq 0$,
\begin{eqnarray}
\|\xi_u(x,t) - \xi_u(x',t)\| \leq \beta_u(\|x-x'\|, t), \mbox{and} \nonumber 
\label{eq:df1}
\end{eqnarray}
\item for any $t$, ${\lim_{\|x -x'\| \rightarrow  0^{+}} \beta_u(\|x-x'\|,t) =  0}$.
\end{enumerate}
\end{defn}
Definition \ref{def:pdf} generalizes the discrepancy functions defined in~\cite{chuchuATVA,duggirala2013verification}.
%According to the definition of discrepancy functions, for system~\eqref{eq:system} with input $u(t)$, 
Observe that at any time $t$, the ball with radius $\beta_u(\delta,t)$ centered at $\xi_u(x_0,t)$  contains the reach set of~\eqref{eq:system} starting from $B_{\delta}(x_0)$. Therefore, by bloating the simulation trajectories
$\xi_u(\cdot)$ using the corresponding discrepancy function, we can obtain reach set over-approximations. 
%Similar ideas have been considered based on abstraction techniques to synthesize controllers \cite{zamani2012symbolic}.
%Definition \ref{def:pdf} corresponds to the definition of discrepancy function (Definition 2) in \cite{DMVemsoft2013}, except that we allow an arbitrary $M$-norm as a metric. 
%Note that here we do not require that trajectories converge to each other.
Several techniques for computing discrepancy functions for closed systems have been presented in the literature (see~\cite{chuchuATVA} and the references therein).
The technique discussed  next works for open systems and  exploits the fact that the input signal is fixed for  trajectories starting from different initial states.

%We will assume that the function $f$ is also continuously differentiable. 
%\sayan{Introduce assumptions up front.}
%\subsection{Computing discrepancy from Jacobian matrices}

First, we introduce a basic result that follows from the high-order mean value theorem (Lemma~\ref{lemma:highdimensionmean}) to connect the differential equation with its Jacobian matrices. Then we show that the terms of the Jacobian matrix with respect to the state variables are bounded over compact sets (Lemma~\ref{lemma:ExistInterval}). Using these two results, we establish that the distance between neighboring trajectories actually follows a differential equation related to the bound of the Jacobian matrix (Lemma~\ref{lemma:distance-ode}). Finally, we prove that the upper bound on the largest eigenvalue of the symmetric part of the Jacobian provides us with a suitable discrepancy function (Theorem~\ref{lemma:final}).

The {\em Jacobian} of $f$ with respect to the state $J_x$ and the input $J_u$ are matrix-valued functions of all the first-order partial derivatives of $f$: 
% with respect to the state components. 
% Similarly, the Jacobian of $f$ with respect to the input, $J_u(x,u)$, is a $n \times m$ matrix-valued function:
\begin{align*}
 \left (J_x(x,u) \right )_{ij} =\frac{\partial f_i(x,u)}{\partial x_j}; ~~~
 \left (J_u(x,u) \right )_{ij}=\frac{\partial f_i(x,u)}{\partial u_j}.
\end{align*}
%
%The Jacobian matrices for Example~\ref{example:running} are: 
%\begin{align*}
%\scriptsize
%J_x(x,u) = \left[ \begin{array}{cc} -3x_1^2 - 1.8x_1 - 0.9 &  2u \\ 1 & -2  \end{array}\right];
%J_u(x,u) =\left[ \begin{array}{c} 2x_2 \\ 0  \end{array}\right].
%\end{align*}
%
%
%We further assume that the function $f$ of \eqref{eq:system} has continuous first-order partial derivatives w.r.t. $x$.
%% and its partial derivatives with respect to $x$ are continuous. 
%%
%The {\em Jacobian} of $f$, ${J}_f: \reals^n \times \reals^l \rightarrow \reals^{n \times n}$, is the matrix-valued function of all the first-order partial derivatives of $f$ with respect to $x$. Let $f_i : \reals^n \times \reals^l \rightarrow \reals$ be the scalar components of $f$, for $i=1\dots n$. The $i^{\textrm{th}}$ row, $j^\textrm{th}$ column of $J_f$ is $\left [J_{f}(x,p) \right ]_{ij} = \frac{\partial f_i(x,p)}{\partial x_j}$. 
%
%
The following lemma from \cite{chuchuATVA} relates $f$ with its Jacobian matrices based on the generalized mean value theorem, see~\cite{chuchuATVA} for the detailed proof.
%\sayan{Either cite a textbook for the generalized mean-value theorem, or say that the details of the proof in ATVA if that is the case. Shouldn't $w$ also be quantified in the following?}
%Next we illustrate a lemma which relate the vector field $f$ with its Jacobian matrix $J_f$. It is the generalized mean value function introduced in \cite{Fan15}.

%\sayan{This would also be a good time to introduce the example.}

\begin{lem} 
\label{lemma:highdimensionmean} 
For any continuously differentiable vector-valued function $f:\reals^{n} \times \reals^{m} \rightarrow \reals^{n}$, $x,r \in \reals^{n}$ and $u,w \in \reals^{m}$,
\begin{equation}\label{eq:high-mean}
\begin{array}{l}
f(x+r,u+w)-f(x,u) = \\
\left(\int_{0}^{1}{J_x(x+s r, u+w )ds} \right) \cdot r + \left(\int_{0}^{1}{J_u(x, u+\tau w )d\tau} \right) \cdot w,
\end{array}
\end{equation}
where the integral is component-wise.
\end{lem}

If $f$ is continuously differentiable, all terms in the Jacobian matrix are continuous. Since the input signals are bounded, i.e., $\forall t >0, u(t) \in \U \subset \reals^m$, the Jacobian matrix $J_x(x,u)$ over compact sets is also bounded:
%\sayan{You want the terms of the Jacobian to be continuous functions. How does continuous differentiability give this?}
%\sayan{Dont throw in new assumptions late? Introduce the necessary assumptions when you first introduced $u$, does not seem like these are strong assumptions. Restate the next lemma accordingly.}
%\sayan{Interval matrices not introduced so far.}
%\sayan{$J_u$ is not used in the lemma.}
\begin{lem}\label{lemma:ExistInterval}
%If the Jacobian matrix of $f$ in system~\eqref{eq:system} is $J_x(x,u)$, then f
For any compact sets $S$, $\U$ 
there exists an interval matrix $\interval [\underline{A},\bar{A}] $ such that
\[
 \forall x \in S, u \in \U,
J_x (x,u) \in \interval [\underline{A},\bar{A}] .
\]
\end{lem}
The lemma follows from the fact that each term of $J_x(x,u)$ is a continuous function of $x,u$, and over the compact domains $S,\U$, the function has a maximum and minimum value that defines the matrix pair $[\underline{A},\bar{A}]$. The bounds of such values can be computed for a broad class of nonlinear functions using   optimization and interval arithmetic solvers.

\begin{lem} \label{lemma:distance-ode}
Fix an input signal $u(t)$.
% and an initial set $\Theta$ for system~\eqref{eq:system}.
Suppose  there exists a compact convex set $S \subseteq \reals^n$ and a time interval $[0,t_1]$ such that for any $x \in \Theta$, $\forall t \in [0,t_1]$, $\xi_u(x,t) \in S$.
%\sayan{for any x on the line connecting x and x'? Why "for any x" and $\forall \ t$ ?}
%\item  $\U \subseteq \reals^m$ is a bounded set such that $\forall t \in [0,t_1], u(t) \in \U$,
%\sayan{Doesn't $u$ by definition have the range $\U$? See para after Eq (1).}
%\item there exists an interval matrix $\interval [\underline{A},\bar{A}] $, such that $\forall x \in S, u \in \U, J_x(x,u) \in \interval [\underline{A},\bar{A}] $,
%\sayan{Again, this does not seem like a new assumption. You already showed that such an interval matrix exists.}
Then for any $x,x' \in \Theta$, for any fixed $t \in [0,t_1]$, the distance $y_u(t) = \xi_u(x',t) - \xi_u(x,t)$ satisfies $\dot y_u(t) =  {A(t)} y_u(t)$, for some $ {A(t)} \in \interval [\underline{A},\bar{A}] $, where $\interval [\underline{A},\bar{A}]$ is an interval matrix satisfying 
Lemma~\ref{lemma:ExistInterval}.
%such that $\forall x \in S, u \in \U, J_x(x,u) \in \interval [\underline{A},\bar{A}] $.
\end{lem}
%Lemma~\ref{lemma:distance-ode} is  proved by differentiating $y_u(t)$ using Lemma~\ref{lemma:highdimensionmean}. %The detailed proof can be found at Appendix \ref{append:proof}.
\begin{pf}
Using Lemma~\ref{lemma:highdimensionmean} we have the following:
\begin{eqnarray} \label{eqn:Delta-diff}
\dot y_u(t) &=& \dot \xi_u(x',t)-\dot \xi_u(x,t)  \nonumber \\
&=& f(\xi_u(x',t),u(t))-f(\xi_u(x,t),u(t)) \nonumber \\
&=& \left( \int_{0}^{1}{J_x(\xi_u(x,t)+ sy_u(t),u(t))ds}\right) \cdot y_u(t) \nonumber \\
&+ & \left( \int_{0}^{1}{J_u(\xi_u(x,t), u(t))ds}\right) \cdot (u(t) - u(t)) \nonumber \\ 
&=& \left( \int_{0}^{1}{J_x(\xi_u(x,t)+ sy_u(t),u(t))ds}\right)\cdot y_u(t)
\end{eqnarray}
%where $y_u(t)$ is the distance $\xi_u(x',t)-\xi_u(x,t)$ starting from $x,x' \in S$.

Given a compact convex set $S$ and bounded set $\U$, 
%\ulrich{You need $\U$ to be compact for Lemma \ref{lemma:ExistInterval}, not just bounded, right?} 
the interval matrix $\interval [\underline{A},\bar{A}] $ satisfies the conditions of 
Lemma~\ref{lemma:ExistInterval}.
For any fixed $t$, 
$\int_{0}^{1}{J_x(\xi_u(x,t)+ sy_u(t),u(t))ds}$
is a constant matrix. Because $\xi_u(x,t),\xi_u(x',t)$ are contained in the convex set $S$, according to the convexity assumption of $S$, $\forall s \in [0,1], \xi_u(x,t)+ sy_u(t)$ is also contained in $S$.  Thus, at $t$, $J_x(\xi_u(x,t)+ sy_u(t),u(t)) \in \interval [\underline{A},\bar{A}] $. Since the integration is from 0 to 1, it is straightforward to check that also
\[ \int_{0}^{1}{J_x(\xi_u(x,t)+ sy_u(t),u(t))ds} \in \interval [\underline{A},\bar{A}] .\] 
We rewrite~\eqref{eqn:Delta-diff} as 
\begin{eqnarray} \label{eqn:ydotset}
\dot y_u(t) =  {A(t)} y_u(t),  {A(t)} \in \interval [\underline{A},\bar{A}], 
\end{eqnarray}
which means that at any fixed time $t \in [0,t_1]$, we always have $\dot y_u(t) =  A(t) y_u(t)$, where $A(t)$ is unknown but $A(t) \in \interval [\underline{A},\bar{A}] $. \qed
\end{pf}
Using the differential equation in Lemma~\ref{lemma:distance-ode}, we can get a discrepancy function by bounding the eigenvalues of $\interval [\underline{A},\bar{A}] $:

\begin{thm}
\label{lemma:final}
Fix the input signal $u(t)$ for system~\eqref{eq:system}. Suppose the assumptions in Lemma~\ref{lemma:distance-ode} hold, and
$\exists \gamma \in \reals$ such that $\forall A(t) \in \interval [\underline{A},\bar{A}], $
\begin{equation}\label{eq:largest_eig_assumption}
eig({A^{T}(t)+A(t)})/2 \leq \gamma; 
\end{equation}
%where $eig()$ means the eigenvalue;
then for any $x, x' \in \Theta$ and for any $t \in [0,t_1]$, $$\| \xi_u(x ,t) - \xi_u(x',t) \| \leq \|x-x'\|e^{\gamma t}.$$
\end{thm}
\begin{pf}
Fixing the two initial states $x,x' \in \Theta$, from  Lemma~\ref{lemma:distance-ode}, we know that $\dot y_u(t) =  {A(t)} y_u(t)$, for some $ {A(t)} \in \interval [\underline{A},\bar{A}] $,  where $y_u(t)=\xi_u(x',t)-\xi_u(x,t)$.
By differentiating $\|y_u(t)\|^2$, we have that for any fixed $t\in[0,t_1]$,
\begin{equation}\label{eq:diff-interval}
\begin{array}{rcl}
\dfrac{d\|y_u(t)\|^2}{d t} & = &  \dot{y}_u^T(t) y_u(t) + {y}_u^T(t)  \dot{y}_u(t) \\
&=& y_u^T(t) \left ( {A(t)}^T + {A(t)} \right ) y_u(t).
\end{array}
\end{equation}
If $eig({A^{T}(t)+A(t)})/2 \leq \gamma$, then the eigenvalues of $B=A^T(t) +A(t) - 2\gamma I$, where $I$ is the identity matrix, are all less or equal
to zero, so $B$ is negative semi-definit. Therefore, \\
%\[
$y_u^T(t) \left ( {A(t)}^T + {A(t)} - 2\gamma I \right ) y_u(t) \leq 0$.
%\]
Hence, \eqref{eq:diff-interval} becomes \\
$
\begin{array}{c}
\frac{d\|y_u(t)\|^2}{d t} \leq 2\gamma \|y_u(t)\|^2.
\end{array}
$
After applying Gr{\"o}nwall's inequality, we have 
$\|y_u(t)\| \leq \|y_u(0)\| e^{{\gamma} t}, \forall t \in [0,t_1].$  \qed
\end{pf}
%Given any matrix $M \succ 0$, $\|y_u(t)\|^2_M = y^T(t)My_u(t)$, and 

%We write $A(t)$ as $A$ in the following for simplicity.
%If there exists a $\gammahat$ such that $$A^TM +MA \preceq \gammahat M, \forall A \in \A,$$
%
%$\frac{\gammahat}{2}$ can also be seen as an upper bound of the matrix measure of the family of matrices $\A$ (see Section \ref{sec:mm}). Since $\mu_M(A) \leq \frac{\gammahat}{2}, \forall A \in \A$ means $ CAC^{-1}+(CAC^{-1})^T \preceq \gammahat I, \forall A \in \A$, where $M = C^TC$. Pre- and post-multiplying the inequality by $C^T$ and $C$, we can also get $A^TM +MA \preceq \gammahat M, \forall A \in \A$.

Theorem~\ref{lemma:final} obviously provides a discrepancy function $$\beta_u (\|x-x'\|,t)=\|x-x'\| e^{{\gamma} t}.$$ However, computing one $\gamma$ for the entire time horizon is usually too conservative to be directly used.
Algorithm 2 in~\cite{chuchuATVA} provides a method for constructing a reachtube
%compute such upper bound $\gamma$ of the eigenvalues for an interval matrix $[\underline{A},\bar{A}]$ using $O(n^2)$ time. 
using one simulation trajectory and initial partition size $\delta$ as input, and produces a sequence of coefficients 
that defines the piece-wise exponential discrepancy function.
The algorithm consists of the following steps:
\begin{inparaenum}[a)]
\item First, using the Lipschitz constant, a coarse over-approximation of the reachable set up to a  short time horizon $T_s$ is constructed. Let this set be $S$. 
\item Compute the  interval matrix $[\underline{A},\bar{A}]$, which bounds the possible values of the Jacobian matrix $J_{x}(x,u)$.
\item Compute the largest eigenvalue $eig\left((\underline{A}+\bar{A})+(\underline{A}+\bar{A})^T\right)/2$. From this value, an upper bound $\gamma$ of the eigenvalue of $(A+A^T)/2$ for all $A \in [\underline{A},\bar{A}]$ is computed using a theorem from matrix perturbation theory.
\item The upper bound $\gamma$ (possibly negative) defines the discrepancy function $\beta_u(\delta,t) = \beta_u'(\delta,t_0)e^{\gamma(t-t_0)}$ over the simulation time interval $[t_0, t_0 + T_s]$, where $\beta_u'(\cdot,\cdot)$ is the previous piece of the discrepancy function. Using this piece-wise discrepancy function, an over-approximation of the reachable set is finally computed.
\end{inparaenum} 

\begin{exmp}
For Example~\ref{example:running}, restricting $x_1$ to be within the range $[0.4,0.6]$ and $u$ to be within the range $[0.1,0.2]$ provides $J_x \in \interval[\underline{A},\bar{A}]$ for $\underline{A} =  \left[ \begin{array}{cc} -3.06 &  0.2 \\ 1 & -2  \end{array}\right]$ and $\bar{A} = \left[ \begin{array}{cc} -2.1 &  0.4 \\ 1 & -2  \end{array}\right]$. Using Algorithm 2 in~\cite{chuchuATVA}, we get that $\gamma = -1.05$ satisfies Equation~\eqref{eq:largest_eig_assumption}. Therefore, $\beta_u (\|x-x'\|,t)=\|x-x'\| e^{{-1.05} t}$ is a discrepancy function for this system with fixed input $u(t)$.
% with fixed input $u(t)$ and $\forall t, u(t) \in [-0.1,0.1] $.
\end{exmp}

\section{Verifying system with fixed inputs}
\label{sec:verification_algorithm}

%Although it is generally difficult to get the closed-form solution of dynamic systems, validated simulation libraries, such as VNODE-LP \cite{vnode2006} and CAPD \cite{capd}, use numerical integration to generate a sequence of states with guaranteed error bounds.
%{These simulators} will return a sequence of hyper-rectangles that is guaranteed to contain the solution, 
To implement our novel method for computing discrepancy
functions of open systems, the algorithm for simulation-driven
verification (see Algorithm~\ref{alg:verification}) published
in~\cite{chuchuATVA} can be used with minor
modifications. For sake of completeness, we briefly discuss the key features
here; for more details the reader is referred to~\citep{chuchuATVA}.
%When implementing the above concepts, the representation of the trajectories are simulations, and the representation of the reach sets are reachtubes as defined below.
Throughout this section, we fix an input signal $u(t)$ for the system~\eqref{eq:system}.
%\subsection{From simulations to reachable sets using discrepancy}
%\label{prelim:reachset}

\begin{algorithm2e}[h!]
%	\scriptsize
	\caption{Verification of systems with input}
	\label{alg:verification}
	\SetKwInOut{Input}{input}
%	%%
	\Input{$\Theta,u(t),T,\unsafe,\epsilon_0,\tau_0$}
	$\delta \gets dia(\Theta); \epsilon\gets \epsilon_0; \tau \gets \tau_0; \TB\gets \emptyset$\;
	$\C   \gets \mathit{Cover}(\Theta,\delta,\epsilon)$\;   \label{al1:partition}
	\While{ $\C \neq \emptyset $ \label{al1:while}
	}{ 	
		\For{$\langle x,\delta,\epsilon \rangle\in \C$}{
			$\psi = \{(R_i,t_i)_{i=0}^k\} \gets \mathit{Simulate}(x, u, T, \epsilon,\tau)$\; \label{al1:simulation}
			%		$\beta \gets$ $\findldf$($\psi$,$J_f$,$L_f$,$\delta,\epsilon$)\; \label{al1:discrepancy}
			$\R \gets \mathit{Bloat}(\psi,\delta,\epsilon$)\; \label{al1:bloat}
			\uIf{$\R \cap \unsafe =\emptyset$}{            \label{al1:if}
				$\C\gets \C\backslash\{\langle x,\delta,\epsilon \rangle\}$ \;
				$\TB \gets \TB \cup \R$ \; \label{al1:unsafe}
			}
			\uElseIf{$\exists j, R_j \subseteq \unsafe$}{
				\Return $(\mbox{UNSAFE},\psi)$
			}
			\Else{
				$ \C \gets \C \cup \mathit{Cover}(B_\delta(x),\frac{\delta}{2},\frac{\epsilon}{2}), \tau \gets \frac{\tau}{2}$ \;
			}		\label{al1:endif}
		} 
	} \label{al1:endwhile}
	\Return $(\mbox{SAFE},\TB)$\;
\end{algorithm2e}

Function $\mathit{Cover}()$ returns a set of triples $\{ \langle x, \delta, \epsilon \rangle\}$, where $x$'s are sample states, the union of $B_\delta(x)$ covers $\Theta$ completely, and $\epsilon$ is the precision of the simulation.
Function $\mathit{Bloat}()$ expands the simulation trace $\psi$ by $\beta_u$ to get the reachtube $\R = \{(O_i,t_i)\}_{i=1}^k$. That is, for each $i = 1,\dots,k, O_i \gets \chull{(R_{i-1},R_{i})} \oplus \max_{t \in [t_{i-1}, t_{i}]} \beta_u((\delta+\epsilon),t)$. From Theorem~\ref{lemma:final}, it follows that $\mathit{Bloat}(\psi, \delta, \epsilon)$ contains $\Reach_u(B_\delta(x),[0,T])$.
%and returns a reachtube that contains all the trajectories starting from the $\delta$-neighborhood of $x$.
%
There are two important data structures used in Algorithm~\ref{alg:verification}: $\C$ is a collection of the triples returned by $\mathit{Cover}()$, which represents the subset of $\Theta$ that has not  yet been proved safe, and $\TB$ that stores the bounded-time reachtube.

Initially, $\C$ contains a singleton $\langle x_0,\delta_0 = dia(\Theta),\epsilon_0 \rangle$, where $\Theta \subseteq B_{\delta_0}(x_0)$ and $\epsilon_0$ is a small positive constant. For each triple $\langle x, \delta, \epsilon \rangle \in \C$, the \textbf{while}-loop from Line~\ref{al1:while} checks the safety of the reachtube from $B_{\delta}(x)$, which is computed in Line~\ref{al1:simulation}-\ref{al1:bloat}. $\psi$ is a $(\epsilon,\tau,T)$-simulation from $x$ with input $u(t)$, which is a sequence of time-stamped rectangles $\{(R_i,t_i)\}_{i=0}^k$ and is guaranteed to contain the trajectory $\xi(x,T)$.
% by Definition~\ref{def:simulation}. 
Bloating the simulation result $\psi$ by the discrepancy function $\beta_u$ we get $\R$, a $(B_{\delta}(x),T)$-reachtube with input $u(t)$.
% we have an over-approximation of $\Reach{(B_\delta(x),[0,T])}$. 
%The core function $Bloat()$ will be discussed in detail in the next chapter.
If $\R$ is disjoint from $\unsafe$, then the reachtube from $B_\delta(x)$ is safe and the corresponding triple can be safely removed from $\C$. 
If for some $j$, $R_j$ (one rectangle of the simulation) is completely contained in the unsafe set, then we can get a counterexample of a trajectory that violates the safety property.
Otherwise, the safety of $\Reach_u{(B_\delta(x),[0,T])}$ is inconclusive and a refinement of $B_{\delta}{(x)}$ is made with some smaller $\delta$ and smaller $\epsilon,\tau$.

Recall that the safety verification problem requires us to check whether $\Reach_{u}(\Theta,[0,T])  \cap \unsafe = \emptyset$.  If there exists some $\epsilon>0$ such that $B_{\epsilon}(\Reach_u(\Theta,[0,T]) ) \cap \unsafe = \emptyset$, we say the system is \emph{robustly safe}. 
%That is, the system is robustly safe if all states in some envelope around the system behaviors are safe. 
If there exists some $\epsilon>0, x \in \Theta,$ such that $B_{\epsilon}(R_i) \subseteq \unsafe$ for some $R_i$ in the simulation from $x$, $\{(R_i,t_i)\}_{i=0}^k$, we say the system is \emph{robustly unsafe}.
%
%\sayan{Lets give this algorith as pseudocode.}
%Given the computed discrepancy function $\beta_u$ for the system~\eqref{eq:system}, the safety verification algorithm for~\eqref{eq:system} is provided as Algorithm \ref{alg:verification} in Appendix \ref{append:algorithm}. 
The algorithm returns $\mbox{SAFE}$ if  $\Reach_u(\Theta,[0,T])$ has no intersection with $\unsafe$, along with a robustly safe reachtube $\TB$. It returns $\mbox{UNSAFE}$ upon finding a counter-example, i.e., the simulation $\psi$ with an interval fully contained in $\unsafe$. 

According to Theorem~\ref{lemma:final}, if $\delta$ gets smaller, the value of the discrepancy function $\beta_u$ becomes smaller (i.e., the reachtube is arbitrary close to the simulation), which guarantees that the algorithm always terminates. 
%In other words, the discrepancy function computed using Theorem~\ref{lemma:final} gives us two key properties of the algorithm~\cite{duggirala2013verification}:

\begin{thm} (Soundness \& completeness). Given an unsafe
set $\unsafe$, time bound $T$ and fixed input $u(t)$ for system~\eqref{eq:system}, if Algorithm~\ref{alg:verification} using the discrepancy of Theorem~\ref{lemma:final} returns $\mbox{SAFE}$  or $\mbox{UNSAFE}$, then~\eqref{eq:system} is safe or unsafe, respectively. 
It terminates if ~\eqref{eq:system} is  robustly safe or unsafe.
\end{thm}

When extending this verification algorithm to work for open  hybrid models, the  main complication is that spurious transitions may arise from the  over-approximations in the computed reach sets. Thus,  we have to keep track of possibly spurious mode changes from genuine ones. This is what is implemented in the new version of C2E2 used in Section~\ref{sec:res} for verifying hybrid circuit models.

%% file: models.tex
\section{Modeling of CMOS Circuits}
\label{sec:models}

To investigate the feasibility of our approach, we analyzed models of
\emph{complementary metal-oxide-semiconductor} (CMOS) circuits, the
most common technology nowadays. Its basic components are two types
of transistors (nMOS and pMOS), which can be used to 
build any desired logic. Essentially, both deliver current based on the
voltages applied to their gate (G), drain (D) and source (S) contact. Modern
digital simulation tools like Modelsim or NC-Verilog consider transistors
as simple switches, however. Such tools allow fast functional and timing analysis of 
complex circuits, but lack sufficient accuracy for critical parts of a
circuit design. The latter is provided by analog simulations, using 
state-of-the-art tools like \textit{Spice}. They are capable of 
handling very detailed transistor models, %like Berkeley's \textit{BSIM},
consisting of tens to hundreds of equations and configured by hundreds of
manufacturer-provided 
parameters. However, analog simulations quickly reach their limits in 
terms of simulation complexity for circuits consisting of more than a few tens of transistors and/or signal traces
beyond milliseconds in real-time.

% Most digital circuits today are still manufactured using \emph{complementary 
% metal-oxide-semiconductor} (CMOS) technology, with transistor sizes down to below 20
% nanometers. Modern %design tools provide
% digital simulation tools like Modelsim or NC-Verilog allow fast
% functional and timing analysis of complex circuits consisting of
% millions of transistors. Whereas the used delay models
% provide sufficient accuracy for many applications,
% critical parts of a circuit require a more careful analysis using analog
% simulations. A prominent tool for this purpose is \textit{Spice}, as it handles
% very detailed transistor models, %like Berkeley's \textit{BSIM},
% configured by hundreds of manufacturer-provided 
% parameters. However, these models quickly reach their limits in terms of simulation complexity 
% for circuits consisting of more than a few tens of transistors and/or signal traces
% beyond milliseconds in real-time.

In order to decrease simulation times, simplified models have been
developed~\citep[e.g.][]{arora1993}. They are  smaller 
than \textit{Spice} models (at most six equations are required),
and thus amenable to general simulation tools like~\textit{MATLAB}. Despite the reduced 
complexity, these models can still capture subtle phenomena like
channel length modulation and carrier velocity saturation.

% In order to decrease simulation times, simplified models of CMOS 
% transistors have been introduced~\cite{arora1993}. Their size is significantly
% smaller compared to \textit{Spice} models (at most six equations are required), which makes it
% possible to evaluate them using general 
% purpose tools like \textit{MATLAB}. Despite their reduced 
% complexity these models can capture subtle phenomena like
% channel length modulation (CLM) and carrier velocity saturation.

\begin{figure}[htbp!]

\begin{minipage}[b]{3.2cm}

%   \begin{subfigure}{0.45\linewidth}
%   \centering
%   \scalebox{0.75}{
%   \begin{tikzpicture}
%     \draw
%     (0,0) node[nfet] (NMOS) {}
%     (NMOS.D) node[anchor=north west] {$D$}
%     (NMOS.S) node[anchor=south west] {$S$}
%     (NMOS.G) node[anchor=north east] {$G$}
%     ;
%   \end{tikzpicture}
% }
% %  \caption{\footnotesize NMOS}
%   \end{subfigure}
% \\
% \\
% \\
%   \begin{subfigure}{0.45\linewidth}
%   \centering
%   \scalebox{0.75}{
%   \begin{tikzpicture}
%     \draw
%     (0,0) node[pfet,yscale=-1] (PMOS) {}
%     (PMOS.D) node[anchor=south west] {$D$}
%     (PMOS.S) node[anchor=north west] {$S$}
%     (PMOS.G) node[anchor=south east] {$G$}
%     ;
%   \end{tikzpicture}
% }
%  % \caption{\footnotesize PMOS}
%   \end{subfigure}
  \scalebox{0.82}{
  \input{inverter.tex}}
  \caption{\small Internal structure of CMOS inverter}
  \label{fig:CMOSInverter}
%\end{figure}
\end{minipage}
~~~
\hspace{0.2cm}
% \begin{minipage}[b]{0.01cm}
% \end{minipage}
~~~
\begin{minipage}[b]{3.8cm}
  \centering
  \scalebox{0.6}{
  \input{nor.tex}}
  \caption{\small Internal structure of NOR gate}
  \label{fig:nor}
\end{minipage}
\end{figure}

Actually, NMOS and PMOS transistors
%(see Figure~\ref{fig:CMOSTransitorDetail})
differ substantially in physical and, hence, electrical
properties. In our
model~\citep{Mai17}, every
transistor operate in one of three different operation regions:
the sub-threshold (ST) region, where very little
current is delivered, the ohmic region (OHM), where the current scales
linearly, and the saturation region (SAT), where the current only changes
moderately. The actual behavior within every region is described by a set of
differential equations, which involve several fitting parameters. 
These differ for NMOS and PMOS transistors and
are either inferred directly from \textit{Spice} model variables or fitted 
to \textit{Spice} simulations.

% In the (SUB) region the current
% through the transitor $I_D$ was set to $0$, i.e., an open circuit. For the (OHM)
% region the models specifies

% \begin{equation} 
% \label{eq:ID}
% I_D=\frac{\mu_{s}C_{ox}W}{L\big(1+\frac{\mu_{s}V_{D}}{L v_{sat}}\big)}(V_{G}-V_{T}-0.5\alpha V_{D})V_{D}
% \end{equation}

% In the (SAT) region only the channel length modulation is important ending up with

% \begin{equation} 
% I_D=I_{Dsat}\frac{L}{L-l_d}
% \end{equation}

% where $I_{Dsat}$ is $I_D$ for $V_D$ equal to

% \begin{equation} 
% V_{Dsat} = \frac{L v_{sat}}{\mu_{s}}\Big(\sqrt{1+\frac{2\mu_{s}(V_G-V_T)}{\alpha L v_{sat}}}-1\Big)
% \end{equation}

% and

% \begin{equation} 
% l_d = Lln\Big(1+\frac{V_D-V_{Dsat}}{V_P}\Big)
% \end{equation}

\subsection{Hybrid inverter model}
\label{sec:hybrid}

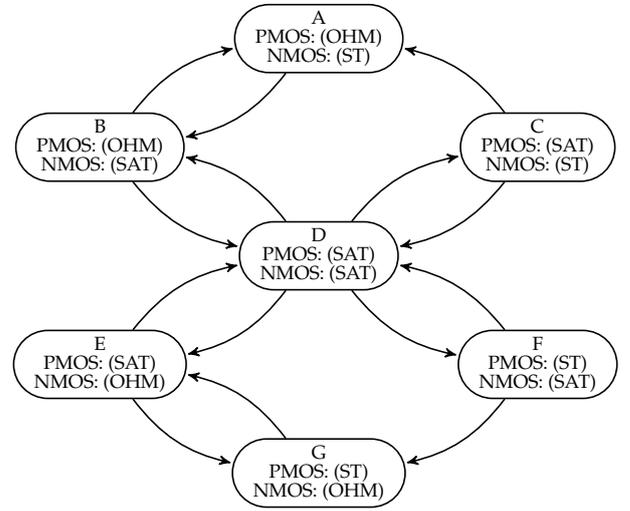
\begin{figure}[tbhp!]
%\begin{subfigure}[b]{0.42\textwidth}
\centering
\tikzset{rectangle state/.style={draw, rounded rectangle},every edge/.append}

\begin{tikzpicture}[->,>=stealth',shorten >=1pt,auto,node distance=2cm, scale = 0.72, semithick, transform shape]

  \node[rectangle state] (A)                    [align=center]{A\\ PMOS: (OHM)\\ NMOS: (ST)};
  \node[rectangle state] (B) [left of=A,left of=A,below of=A] [align=center]{B\\ PMOS: (OHM)\\NMOS: (SAT)};
  \node[rectangle state] (C) [right of=A,right of=A,below of=A]     [align=center]{C\\ PMOS: (SAT)\\NMOS: (ST)};
  \node[rectangle state] (D) [below of=A,below of=A]     [align=center]{D\\ PMOS: (SAT)\\NMOS: (SAT)};
  \node[rectangle state] (E) [left of=D,left of=D,below of=D]     [align=center]{E\\ PMOS: (SAT)\\NMOS: (OHM)};
  \node[rectangle state] (F) [right of=D,right of=D,below of=D]     [align=center]{F\\ PMOS: (ST)\\NMOS: (SAT)};
  \node[rectangle state]         (G) [below of=D,below of=D]     [align=center]{G\\ PMOS: (ST)\\NMOS: (OHM)};
	\path (A) edge [bend left = 20] node [pos=0.5, sloped, above] {} (B)
        (B) edge [bend left = 20] node [pos=0.5, sloped, above] {} (A)
             edge [bend right = 20] node [pos=0.5, sloped, above] {} (D)
        (C) edge [bend right = 20]  node [pos=0.5, sloped, below] {} (A)
            edge [bend left = 20]	  node [pos=0.5, sloped, above] {} (D)
        (D) edge [bend right = 20] node [pos=0.5, sloped, below] {} (B)
           	edge [bend left = 20] node [pos=0.5, sloped, above] {} (C)
				edge [bend left = 20] node [pos=0.5, sloped, below] {} (E)
				edge [bend right = 20] node [pos=0.5, sloped, above] {} (F)
        (E) edge [bend left = 20] node [pos=0.5, sloped, above] {} (D)
            edge [bend right = 20]  node [pos=0.5, sloped, above] {} (G)
        (F) edge [bend right = 20]  node [pos=0.5, sloped, below] {} (D)
            edge [bend left = 20] node [pos=0.5, sloped, below] {} (G)
        (G) edge [bend right = 20] node [pos=0.5, sloped, below] {} (E);
\end{tikzpicture}
\caption{{\small Hybrid model automata of a CMOS inverter. The nodes represent
the different modes, each involving specific transition guards and 
invariants. The label shows the operation regions of the transistors,
the arcs indicate the possible transitions.}}
\label{fig:HybridInverterModel}
\end{figure}

The simplest CMOS gate is an inverter (see
Figure \ref{fig:CMOSInverter}), which consists of two transistors stacked above
each other. Its output voltage $V_{out}$ is the inverse
of the input voltage $V_{in}$, ideally $V_{out}=V_{DD}-V_{in}$ with $V_{DD}$
denoting the supply voltage. In reality, $V_{out}$ is determined by
\begin{equation} \label{eq:main}
\dot{V}_{out}=\frac{1}{C_L}I_{out}
\end{equation}
where $C_L$ is the external load capacitance seen by the output. The output current $I_{out}$
is the difference between the current delivered by \myTrans{1} and the current 
consumed by \myTrans{2}, which both depend upon $V_{in}$ and $V_{out}$. 
Since each of the two transistors can operate in three different regions, 
our basic hybrid inverter model has nine modes. As two of those modes are
unreachable in reality, the hybrid model (called {\InvHy}
shown in Figure~\ref{fig:HybridInverterModel}  
in the sequel) has only $7$ modes.

% The simplest CMOS gate is an inverter (shown in
% x~\ref{appendix:inv} as Figure \ref{fig:CMOS}), which consists of a PMOS transistor stacked above
% an NMOS one. Its output voltage $V_{out}$ is the inverse
% of the input voltage $V_{in}$, ideally $V_{out}=V_{DD}-V_{in}$ where $V_{DD}$
% denotes the supply voltage. In reality, $V_{out}$ is determined by
% \begin{equation} \label{eq:main}
% \dot{V}_{out}=\frac{1}{C_L}I_{out}
% \end{equation}
% where $C_L$ is the external load capacitance seen by the output. The output current $I_{out}$
% is the difference between the current delivered by the PMOS and the current 
% consumed by the NMOS, which both depend upon $V_{in}$ and $V_{out}$. 
% As each of the two transistors can operate in three different regions, 
% our basic hybrid inverter model has nine modes. As two of those modes are
% unreachable in reality, we finally arrive at the hybrid
% model shown in Appendix \ref{appendix:inv} Figure~\ref{fig:CMOS_automata} (called {\InvHy} in the sequel). 

\subsection{Uniform model}
\label{sec:uniform}

Given that the number of modes increases exponentially with the number of
transistors in a circuit, it is natural to
consider ways of avoiding multiple modes already in the transistor models:
If the behavior of a transistor could be described by 
a single, possibly more complex equation that is valid for all operation 
regions, the need for a hybrid model vanishes altogether. 

Our \emph{uniform model} 
{\InvUni}~\citep{Mai17} accomplishes this by smoothening the boundaries 
between different regions, using suitably chosen continuous functions.
This results in a single non-linear equation (involving exponentials and 
logarithms), which describes
the current through the transistor over the whole operation range.
In conjunction with equation~\eqref{eq:main}, this finally leads to a non-linear ODE
that describes the behavior of $V_{out}$ depending on $V_{in}$. 
Empirical validations using Spice simulations revealed a surprisingly
good modeling accuracy. 

%\begin{figure}[ht]
% \centering
%  \scalebox{0.7}{
%  \input{Fig/nor.tex}}
%  \caption{\small Internal structure of NOR gate.}
%  \label{fig:nor}
%\end{figure}

Apart from dramatically reduced model complexity, a key feature of our uniform model 
is the straightforward development of models for multi-transistor circuits like
the {\NOR} gate shown in Figure~\ref{fig:nor}. In a hybrid model,
this gate would blow up to a system of $3^4=81$ states; here, we end up with a 
system of two non-linear ODEs only:
%\begin{equation*}
%\label{eq:nor}
\[
\begin{array}{l}
\dot{V}_{m}=\frac{1}{C_M}(I_{1}-I_{2}); 
\dot{V}_{out}=\frac{1}{C_L}(I_{2}-I_{3}-I_{4})
\end{array}
\]
%\end{equation*}
Herein, $I_{X}$ represents the current through transistor \tikz \node[label, font=\tiny] {$X$};.
The change of $V_m$ is proportional to the current charging $C_M$
(cp.\ eq.~\eqref{eq:main}) and is just the difference between the currents flowing 
through the transistors \myTrans{1} and \myTrans{2}.
Note that $C_M$ represents the capacitances of the transistor contacts only, and
is hence several orders of magnitude smaller than $C_L$. The derivative
of $V_{out}$ is finally determined by the current passing through
\myTrans{2} minus the ones consumed by the transistors
\myTrans{3} and \myTrans{4}.

% Here, $I_{X}$ representing the current through transistor \tikz \node[label, font=\tiny] {$X$};.
% The derivative of $V_m$, i.e., the current flowing to $C_M$ divided by $C_M$,
% compare eq.~\eqref{eq:main}, is just the difference between the current flowing 
% through the PMOS transistors \myTrans{1} and
% \myTrans{2}.
% Note that $C_M$ represents the capacitances of the transistor contacts only, and
% is thus several orders of magnitude smaller than $C_L$. The derivative
% of $V_{out}$ is finally determined by the current passing through the lower PMOS
% \myTrans{2}
% minus the currents consumed by the NMOS transistors
% \myTrans{3} and \myTrans{4}.

%US: Dropped, as it does not reveal much ...
%
% \begin{figure}[t!]
% \centering
% \scalebox{1}{
%  \input{Fig/hybrid_automata.tex}}
% \includegraphics[height=2in]{Fig/CMOS}
% \caption{{\small Hybrid model of CMOS Inverter.}}
% \label{fig:HybridInverterModel}
% \end{figure}

%% file: inverter.tex
\begin{tikzpicture}
  \draw
  (0,0) node[pfet,yscale=-1] (PMOS) {}
  (PMOS.G) node [anchor=south, label] {$1$}
  (PMOS.D) node[nfet,anchor=D] (NMOS) {}
  (NMOS.G) node [anchor=north, label] {$2$}
  (NMOS.S) node[ground] {}
  (NMOS.G) to [short] node[pos=0,circ] (IN) {} (PMOS.G)
  (IN) to [short,*-o] ++ (-0.5,0) node[anchor=east] (VIN) {$V_{in}$}
  (PMOS.D) to[short,*-o] node[circ,pos=0] (CONN) {} ++(1.5,0) node[anchor=west] {$V_{out}$}
  (CONN) to[C=$C_L$] (CONN |- NMOS.S) node[ground] {}
  (PMOS.S) to [short,-o] ++(0,0.5) node[anchor=south] {$V_{DD}$}
  ;
\end{tikzpicture}

%% file: nor.tex
\begin{tikzpicture}[yscale=1, font=\Large]

  \draw
  (0,0) node[pfet,yscale=-1] (P1) {}
  (P1.D) node[pfet, yscale=-1, anchor=S] (P2) {}
  (P1.D) node[anchor=east] {$V_m$}
  (P1.D) to [short,*-] ++(0.1,0) to [C=$C_M$] ++(2,0) node [ground,rotate=90] {}
  (P2.D) node[nfet, anchor=D] (N1) {}
  (N1.S) node[ground] {}
  (P1.S) to [short,-o] ++(0,0.5) node [anchor=south] {$\vdd$}
  (P1.G) node [anchor=south, label] {$1$}
  (P1.G) to ++(-0.5,0) node [anchor=east] {$V_a$}
  (P2.G) to ++(-0.5,0) node [anchor=east] {$V_b$}
  (P2.G) node [anchor=north, label] {$2$}
  (N1.G) node [anchor=north, label] {$3$}
  (N1.G) to ++(-0.5,0) node [anchor=east] {$V_a$}
  (N1.D) to [short,*-*] ++(2.4,0) node[nfet, anchor=D] (N2) {}
  (N2.S) node[ground] {}
  (N2.D) to [short,-*] ++(1,0) node (CONN) {} to [short,-o] ++ (1,0) node  [anchor=west] {$V_{out}$}
  (N2.G) node [anchor=north, label] {$4$}
  (N2.G) to ++(-0.5,0) node [anchor=east] {$V_b$}
  (CONN) to [C=$C_L$] ++(0,-1.6) node [ground] {}
  ;
\end{tikzpicture}

%% file: res.tex
%!TEX root = main.tex
\section{Experiments and Results}
\label{sec:res}

We have implemented the discrepancy computation and verification algorithm for
open models in the new version of C2E2 and used it to verify several challenging
CMOS circuits (see footnote \ref{modellink}).
%\footnote{Files can be found at \url{https://publish.illinois.edu/c2e2-tool/gate/}.}.
%Apart from analog waveform simulations, we used C2E2's key asset, the reach tubes,
%for verification purposes, which even included metastable behavior.
Due to lack of space, we restrict our discussion here  to a few examples that
demonstrate the principal feasibility, as well as particular strengths, of our 
approach. Experimenting with larger and more complex circuits, which is mandatory
for validating scalability, for example, will be part of our future work.

\subsection{Input, simulation and verification}
\begin{figure}[ht!]
  \centering
  \begin{minipage}{0.24\textwidth}
    \includegraphics[width = \textwidth]{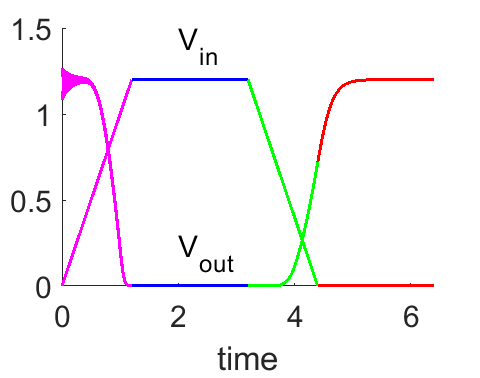}
  \end{minipage}
  \begin{minipage}{0.24\textwidth}
    \includegraphics[width = \textwidth]{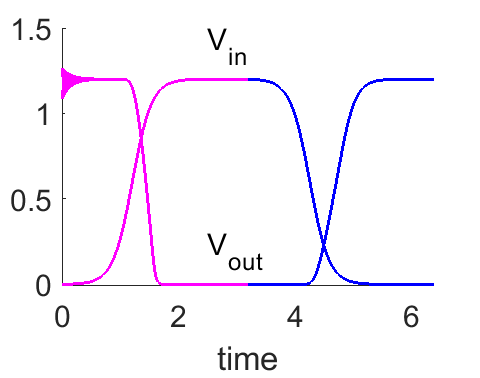}
  \end{minipage}
%  \captionsetup{justification=centering}
  \caption{\small NOR gate output voltage over-approximation set for
    $V_{in}={\ramp}$ (left) and $V_{in}={\sig}$ (right). Different colors
    indicate different modes in the model.} 
  \label{fig:NOR}
\end{figure}

As external input signals,  we use both ramp ({\ramp}) and
sigmoidal signals ({\sig}), which are generated 
using two separate hybrid automata;
a $4$-state one for {\ramp} and $2$-state one for
{\sig}.
We successfully verified several properties of  {\InvHy}, {\InvUni},
{\NOR}-gate, and {\OR}-gate models using the tool. For all the models, 
we set the unsafe set to be $V_{out}>$\,\SI{1.32}{\volt} and the time
horizon to be \myTime{6.4}.
%(parameters see Table~\ref{tab:setting}). 
The first one uses the hybrid model
presented in Section~\ref{sec:hybrid}, so we end up with $7 \times 4=28$ modes
in the {\ramp} case and $7 \times 2=14$ in the {\sig} case. All other circuits
are based on the uniform model. The {\OR} gate is easily
derived from the {\NOR} shown in Figure~\ref{fig:nor} by appending an
inverter. All circuit models based on the uniform model have
very complex descriptions, i.e., hundreds of
logarithmic and exponential terms in their ODEs. 
% and thousands of other arithmetic operations. 
Figure~\ref{fig:NOR} shows some simulation results for the {\NOR}-gate.

%We show the ODEs, verification parameters and reachtube of {\InvHy}, {\InvUni},
%{\NOR} and {\OR} gates with {\ramp} and {\sig} inputs in Appendix
%\ref{appendix:inv}, \ref{append:nor}, \ref{appendix:or}. 

In addition, we also investigated a two-inverter
loop, where the input of one inverter is connected to the output of the
other one, implementing a simple state-holding device. In contrast to the other 
circuits used in our experiments, however, it does not have an
external input. Consequently, we just set the output voltages to 
some initial values and let the circuit run.

Generally, all the simulations behave as expected and show smooth output transitions even when
activated by a ramp at its input. Verification shows that, despite initial
state uncertainty, the unsafe set $V_{out}>$\,\SI{1.32}{\volt} is not reached and simultaneously 
provides the complete reachtube within the safe set.  For example, Fig.~\ref{fig:NOR}
shows that the reachtube converges quickly to a deterministic signal
trace. Total verification time, split between simulation (Sim.) and
discrepancy computation (Discr.), is shown in Table~\ref{tab:result_basic}.

%We will experiment to see the difference in verification result between setting
%$stim$ to be the thin variable and not using thin variable. 
%We failed to successfully verify these models with  other tools for non-linear
%models, specifically, with Flow*\footnote{Flow*'s  parser was not able to handle
%  these large circuit models.}
%and CORA, and therefore are not able to present performance comparisons. 
%
% \footnote{The main limitation of CORA is that the splitting in the case of excessive
% over-approximations may result 
% in an exponential complexity. We have experienced this problem
% trying to analyse the Example 1, leading to a stack overflow caused 
% by reaching the maximum recursion limit.}

%\begin{table}
%\centering
%\begin{center}
%	\captionof{table}{\small Verification parameter settings} \label{tab:setting} 
%    \begin{tabular}{| c | c |}
%    \hline
%	Verification parameters & Setting\\ \hline
%%	Initial Set & $V_{out}\in[1.19V,1.21V],stim=0V,t=0s$\\ \hline
%	Unsafe Set & $V_{out}>1.32V$ \\ \hline
%	Time Horizon & $6.4s$\\ \hline
%    \end{tabular}
%\end{center}
%\end{table}

 \begin{table}[tbhp!]
%\centering
\caption{\small Verification of {\InvHy}, {\InvUni}, {\NOR}-gate and {\OR}-gate with {\ramp} (top) and 
{\sig} (bottom) input and {\InvLoop} without input on a standard laptop (16G RAM, Intel Core i7 CPU). All verification results are safe.}
\label{tab:result_basic}
\def\arraystretch{1.2}
\hspace{-0.2cm}
\scalebox{0.9}{
\begin{tabular}{|c | cc | cc c | c |}
\hline
\mur{Model} & \multicolumn{2}{c|}{ Verification parameters} &
        \multicolumn{3}{c|}{Timing split [s]} & \mur{time [s]} \\
\cline{2-6}
& Steps & Initial Set & Sim. & Discr. & I/O &  \\
\hline
{\InvHy} & 25.6k &  $V_{out} \in [1.15,1.2]$ & 7 & 5 & 2 & 14\\
\hline
{\InvUni} & 12.8k & $V_{out} \in [1.15,1.2]$ & 7 & 12 & 2 &21 \\
\hline
{\NOR} & 320k & $V_{out} \in [1.15,1.2]$ & 223 & 708 & 108 &1039\\
\hline
{\OR} & 320k& \makecell{$V_{nor} \in [1.199,1.201]$ \\ $V_{out} \in [0,0.002]$} & 766 & 1406 & 122 & 2294\\
\hline \hline
{\InvHy} &25.6k  &  $V_{out} \in [1.15,1.2]$ & 7 & 1 & 1 & 9\\
\hline
{\InvUni} & 12.8k & $V_{out} \in [1.15,1.2]$ & 3 & 13 & 2& 18\\
\hline
{\NOR} & 320k  & $V_{out} \in [1.15,1.2]$ & 112& 822 & 66 & 1000\\
\hline
{\OR} &  320k   & \makecell{$V_{nor} \in [1.199,1.201]$ \\ $V_{out} \in [0,0.002]$} & 384 & 1617 & 74 & 2075 \\
\hline \hline
{\InvLoop} & 64k & \makecell{$V_{1} \in [1.0,1.2]$ \\ $V_{2} \in [0.5,0.6]$}  & 18 & 119 & 2 & 139 \\
\hline
\end{tabular}
}
\end{table}

%\vspace{-0.3cm}
\subsection{Metastability analysis}

%It is well known that a
Any bistable digital circuit
%, ranging from a simple storage loop over a flip-flop to complex circuits with internal feedback, 
can be driven into a \emph{metastable} state~\citep{Mar81} in which it may output voltage values in the forbidden
region between $0$ and $1$ or experience very high-frequency oscillations for an arbitrary
time, before it resolves to a proper digital state again.
Verifying whether a circuit may experience metastable behavior is challenging
because it arises in highly nonlinear and sensitive parts of its state
space.

In order to demonstrate the capability of C2E2 to predict metastable
behavior correctly, we use an {\OR}-gate with its output fed back to one
of its inputs. This circuit implements a storage loop, which is capable of memorizing a rising
transition on its second input. It has been shown in~\citep{FNNS15:DATE}
that it can be driven into a metastable state, namely, by an input pulse that 
is shorter than the delay of the feedback loop.

\begin{figure}[t!]
\centering
\includegraphics[width=0.9\linewidth,height=0.6\linewidth]{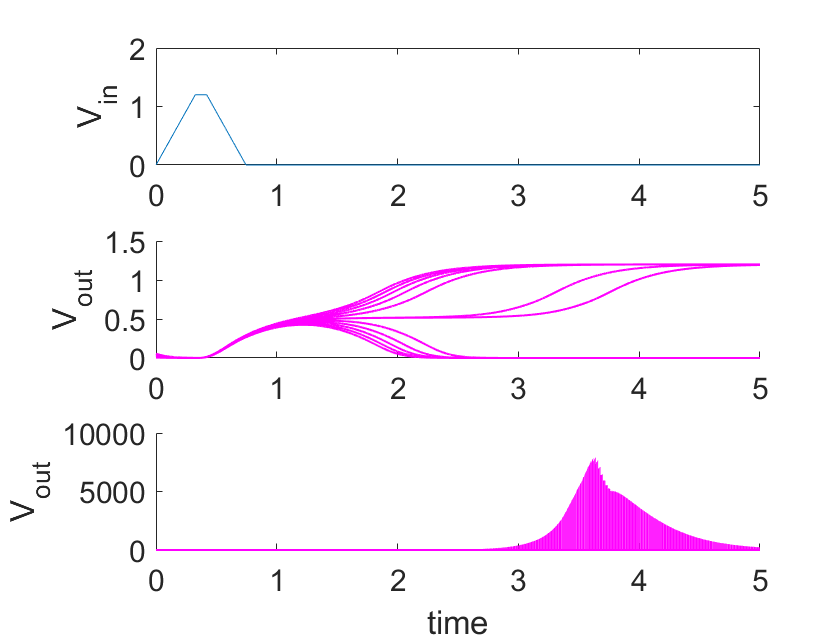}
\caption{\small Metastability analysis of fed back {\OR} gate}
\label{fig:meta}
\end{figure}

Figure~\ref{fig:meta} shows input (top) and simulation traces (middle) of this
circuit computed by C2E2 for different initial values of $V_{out}$. The
reachtube (bottom), corresponding to the output trace sticking longest to a
value around \SI{0.6}{\volt}, shows a blow up to several thousand Volts, which is
physically impossible but indicates the very high sensitivity of the underlying
system of ODEs in the metastable region: Even the slightest disturbances
of the initial state results in very different trajectories, in particular,
in very different metastability resolution times, after which $V_{out}$ 
resolves to $0$ or $1$. Albeit this is in accordance with what 
is known about metastability, to the best of our knowledge, this is the first 
reachability analysis of circuits demonstrating metastable behavior.

%\sayan{Chuchu:
%	This note  came from the last part of section 3. Its rightful place is the para on implementation details or in the conclusion section.
%%Note that in this paper, we are only using the ordinary Euclidean norm for the discrepancy computation. 
%Note that the discrepancy is defined using the Euclidean ($l^2$) norm in Definition~\ref{def:pdf}.
%In~\cite{chuchuEMSOFT}, a less conservative discrepancy function can be achieved by computing the optimal coordinate transformation for the Euclidean norm, which minimizes the upper bound $\gamma$ for the transformed matrix.
%%Such discrepancy function could result in less conservative reachtubes. 
%However, it takes extremely long time to solve such optimization problems for the complicated circuit models in this paper. We thus implemented a simple coordination transformation based on Jordan form decomposition method proposed in~\cite[Sec.~4.3]{chuchuATVA} instead. Empirical results show that the latter method suffices to provide tight reach set over-approximations.
%% and empirical results show that reachtubes can achieve high accuracy.
%%usually converge to single simulation traces after short amount of time.
%}

\subsection{Comparison with existing verification tools}

We provide a comparison with several 
state-of-the-art \emph{nonlinear}\footnote{A comparison with tools like SpaceEx
  and Coho, which support only linear systems, is omitted.} hybrid system
verification tools, namely, \tool{Flow*} \citep{chen2013flow}, %\footnote{\url{flowstar.org}},
\tool{dReach}~\citep{kong2015dreach} %\footnote{\url{dreal.github.io/dReach/}}
and \tool{CORA}~\citep{Althoff2016a}. %\footnote{\url{www.i6.in.tum.de/Main/SoftwareCORA}} 
A comparison of C2E2 with these tools on systems without inputs was previously
reported in~\cite{chuchuATVA, duggirala2013verification}.

\begin{figure*}[t!]
  \centering
  \begin{minipage}{0.29\textwidth}
    \includegraphics[angle=270,width=\textwidth]{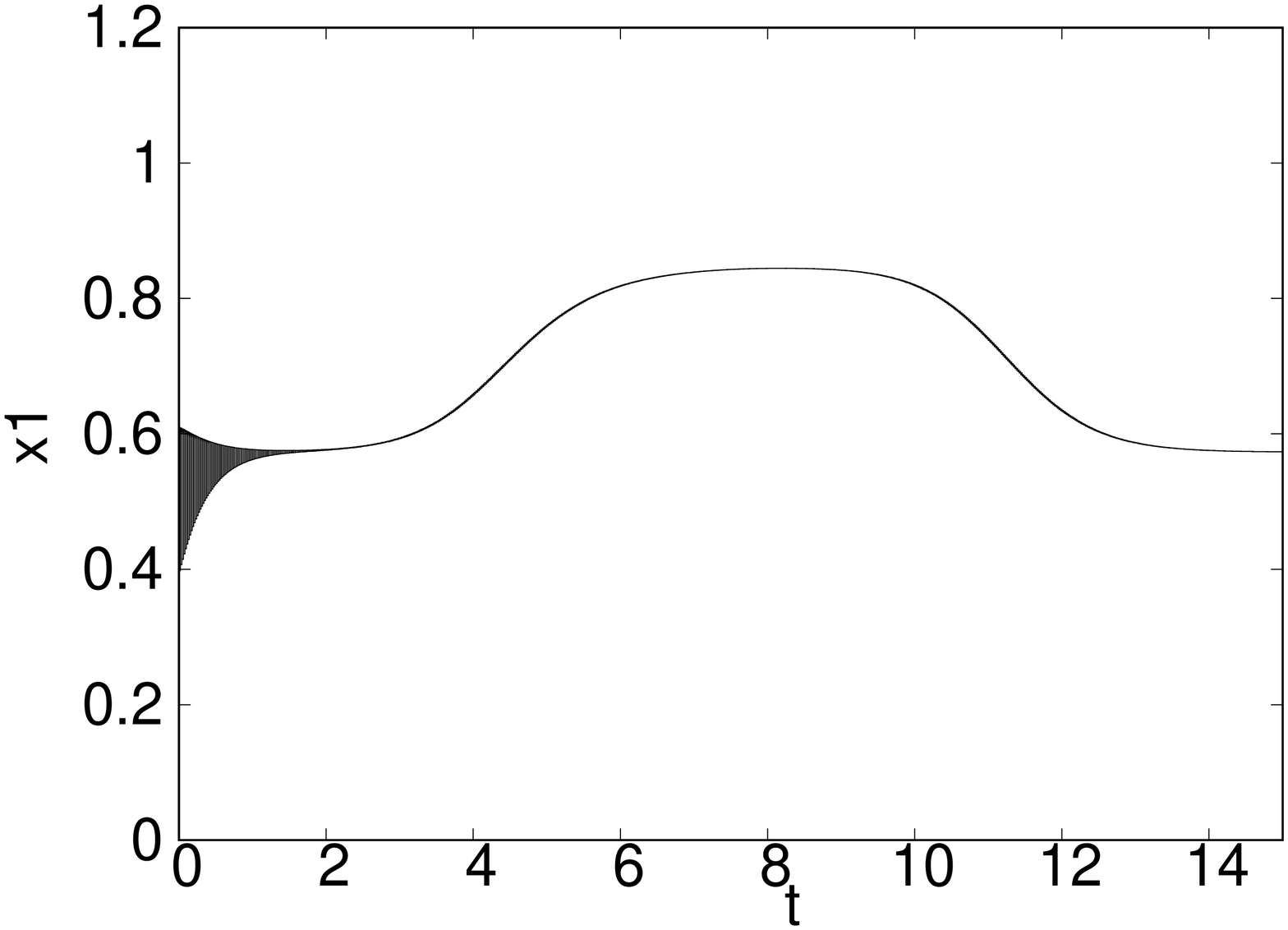}
  \end{minipage}
  \begin{minipage}{0.32\textwidth}
    \includegraphics[width=\textwidth]{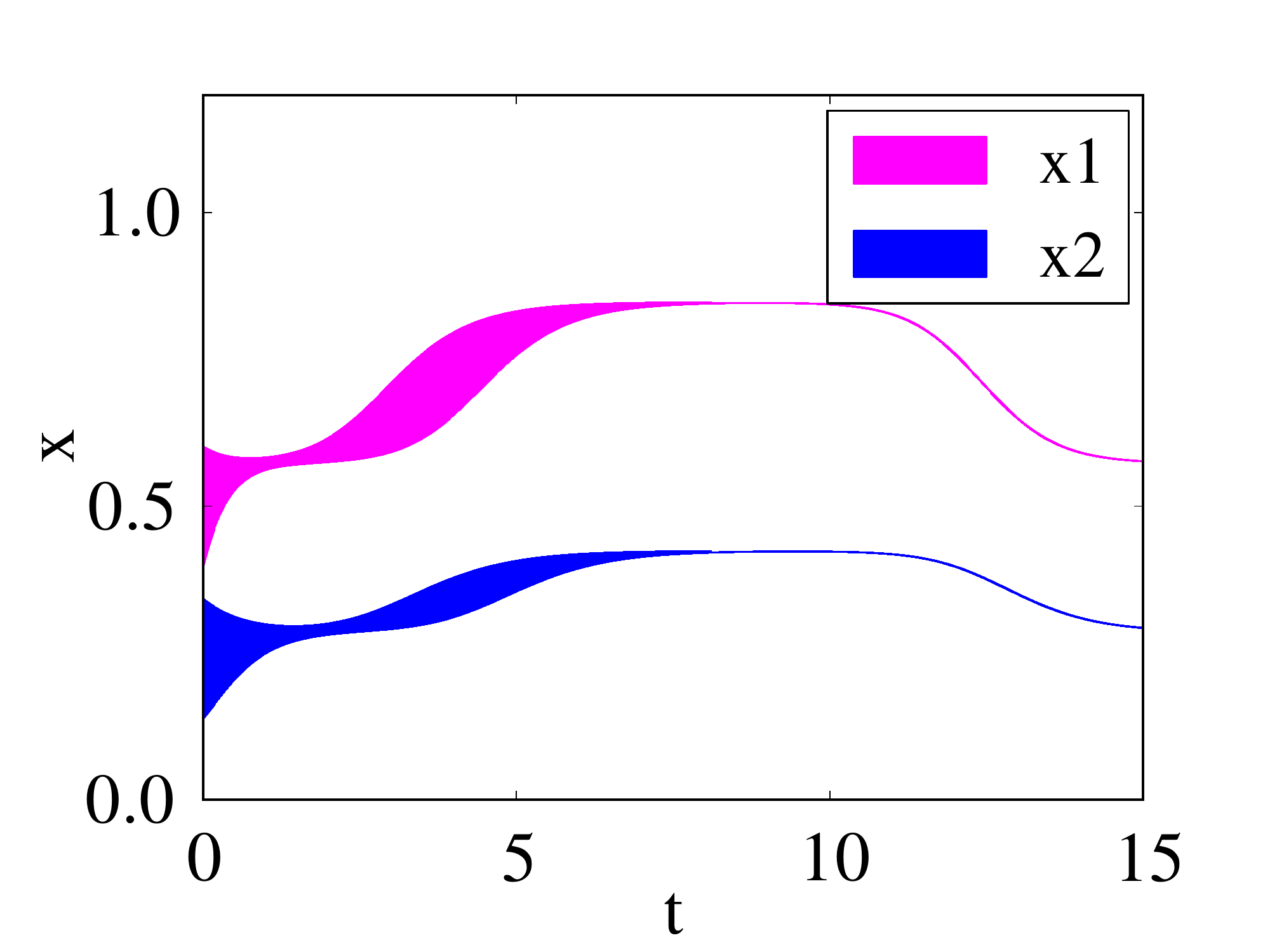}
  \end{minipage}
  \begin{minipage}{0.32\textwidth}
    \includegraphics[width=\textwidth]{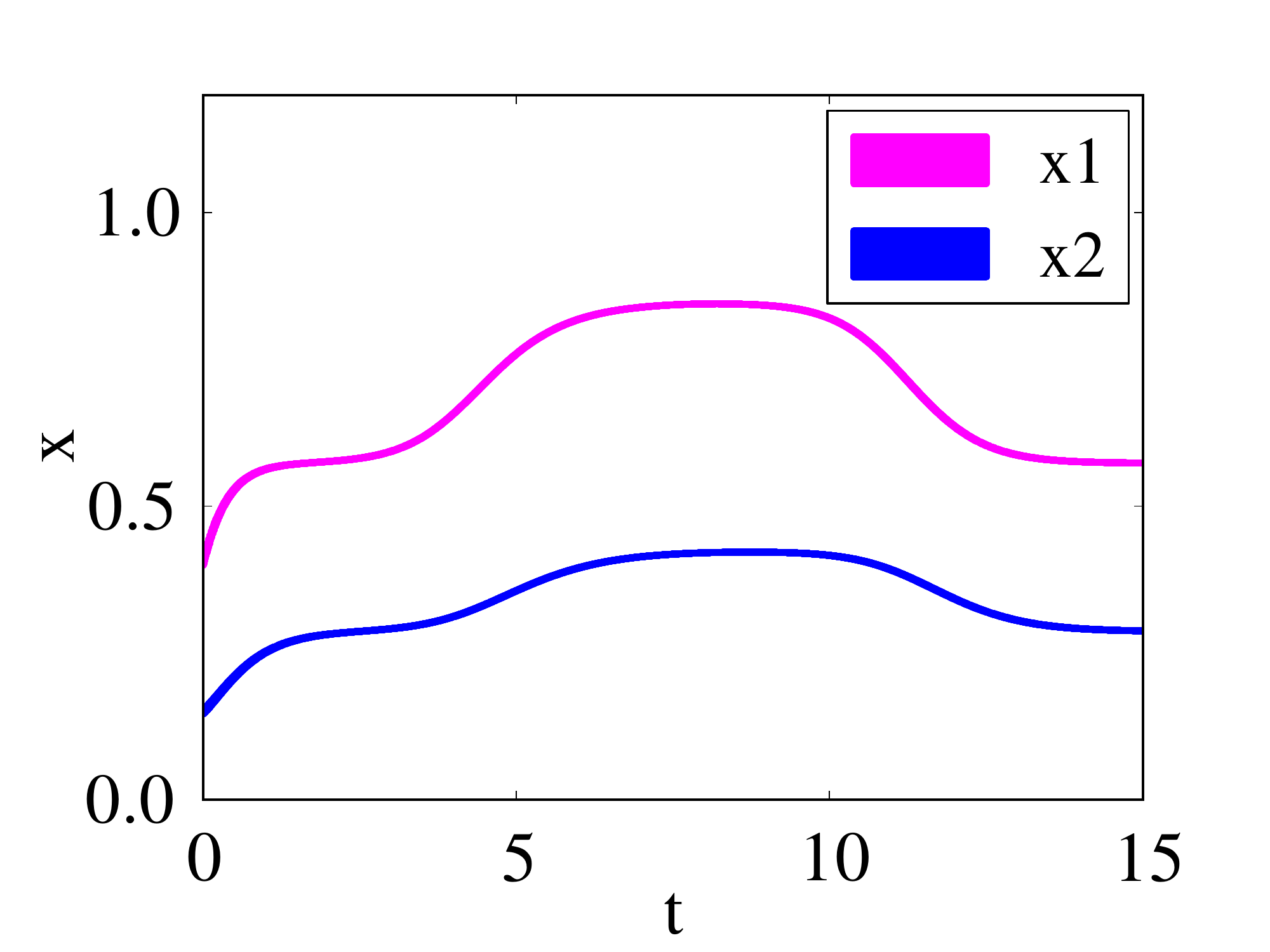}
  \end{minipage}
  % \begin{subfigure}{0.48\linewidth}
  %   \includegraphics[width=\textwidth]{Fig/dReach.png}
  %   \caption{CORA}
  %   \label{fig:Cora}
  % \end{subfigure}
  % ~
  % \begin{subfigure}{0.48\linewidth}
  %   \includegraphics[width=\textwidth]{Fig/dReach.png}
  %   \caption{dReach}
  %   \label{fig:dReach}
  % \end{subfigure}
  \caption{\small Reachability analysis results of cardiac oscillator model (see Example
    \ref{example:running}) in Flow* (left), CORA (middle) and dReach (right).}
  \label{fig:tools}
\end{figure*}

{\bf Flow*} is a reachability-based verification tool for hybrid systems. It
over-estimates the reachable states for nonlinear systems directly on the vector
field and computes Taylor model flowpipes for the ODEs.
While C2E2 performs safety verification for a time horizon of \myTimeS{20} with a
runtime of \myTime{0.1} for the cardiac oscillator in Example~\ref{example:running},
Flow* finishes the flowpipe computation (compare Figure~\ref{fig:tools} (left)) with a runtime of \myTime{5.96}. 
%Flow* terminates with the error ``The remainder estimation is not large enough''
%after \myTimeS{13} had ben simulated (compare Figure~\ref{fig:tools} (left)) which is caused by
%the big aggregation error for the flowpipes over the transition time.
 
We managed to transfer four models from C2E2 to Flow*,
however, none could be verified, for example, on the same set of parameters. {\InvUni}
with ramp input runs for more than $10$ minutes for
the initial mode with the initial set $V_{out} \in [1.18, 1.20]$ and a time step
of \myTime{5e-5} and quits with the error message ``divided by 0'' at a
simulated time of \myTimeS{0.44}. Any changes on these initial parameters (either larger initial set range or time step) resulted in either a ``divided by 0'' or ``segment fault'' error. 
Similar problems occurred with {\InvUni} and sigmoidal input.
% quits with the error ``divided by 0'' after simulating \myTimeS{0.44} with the same initial set and time step. 
%Any variations on the initial set or the time step again led to error messages right from the start. 
With {\InvHy}, the flowpipe computation  keeps oscillating between the modes B and D and the
process was manually killed after $60$ minutes. 
For the remaining models, we were not able to avoid the ``divided by 0'' and the ``segment fault'' errors  by any choice of parameters. We suspect that some of these problems are arising from the large size and complexity of these models. 

{\bf CORA} is a MATLAB toolbox for reachability analysis of 
nonlinear dynamical systems and polynomial differential inclusions~\citep{Althoff2013}.
%. It uses tight non-convex 
%over-approximation of the reachable set of states.
 We  modeled the cardiac oscillator from Example~\ref{example:running}
in CORA as a hybrid 
 system with two modes, with nonlinear dynamics in each, 
 representing the system receiving an upward and a downward 
input stimulus.  Since the initial sets are represented as 
boxes or zonotopes, we needed to also specify a small interval 
(we chose the smallest possible) for the initial conditions 
of the stimulus.  As Figure~\ref{fig:tools} illustrates,  
 the reach set computed by CORA for this example is less precise 
 in the first mode compared to C2E2 (Figure~\ref{fig:simu}).  
 The implementation of our models with CORA was a much 
 harder task than initially anticipated. Only with tremendous support 
 from the author of the tool, who had to 
adapt the original CORA version, we were 
able to run at least the cardiac oscillator and {\InvUni}.
However, the results for these simple examples 
already revealed a quite poor performance compared to
C2E2: For the {\InvUni}, performing the reachability analysis
for a time-horizon of \myTime{0.1} and a time step of \myTime{0.01}
took \SI{222}{} seconds, which caused us to refrain from further experiments
with CORA.

%For the other, more complex models, we were not able to 
%find the right tool parameters to get CORA to work, and always got stuck with 
%some MATLAB errors.
%For this reason we decide to focus 
%our efforts to implement and to compare the analysis of 
%the other more complex models in dReach.

{\bf dReach} is an SMT-based $\delta$-reachability analysis tool for hybrid systems. 
In contrast to C2E2, dReach
picks a single start state in the initial set and attempts to finds a counter-example that
reaches the unsafe set. If it succeeds, a (spurious or real) counter-example has been found 
and the tool finishes. Otherwise, it picks a suitable starting state for the
next trace to be processed. Before it can assure that the goal condition is
unsatisfiable, i.e., the system is safe, the entire initial range has to be covered.
In sharp contrast to C2E2, however, dReach does not display any information
of the reachtubes in that case. To achieve a similar output as provided by
C2E2, one would have to (1) check that no unsafe state is reachable, 
(2) determine the reachtube for each start state to the complement of the
goal condition (i.e., the safe set), and (3) plot their union. 

% measurements by Chuchu
%Cardiac cell, 2.3s, SAT
%Hybrid_inverter_ramp, 13.06s + 20.32s, SAT + SAT
%Hybrid_inverter_sigmoid, 14.26s + 24.4s, SAT + SAT
%Uniform_inverter_ramp, 77.5s, SAT
%Uniform_inverter_sigmoid, 69.9s, SAT

Since the basic operations of dReach and C2E2 are fundamentally different, a fair comparison is 
hard. We managed to build models for the cardiac oscillator in
Example~\ref{example:running} (see Figure~\ref{fig:tools} (right)), {\InvUni}
and {\InvHy}, whereas the latter had to be split 
in up and down transition to work properly. The simulation times, despite the
fact that only a single trace from initial  is computed, is up to
$4 \times$ slower than C2E2.

On the other hand, operationally, dReach is well-suited for metastability analysis,
as bad traces, i.e., ones that end up in the metastable region, are
of interest here. With dReach, this could be achieved by defining the metastable
region as the goal condition. Unfortunately, however, it was not possible
to implement any of the models---{\NOR}, {\OR} or {\InvLoop}---which are the
required  for metastability analysis: The tool issued a 
``log(): domain error'' after short time, which indicates, according to the developers, that
the differential equations are too complex for effectively controlling the errors.

% \begin{table}[htbp!]
% \begin{center}
% \caption{Comparison of different tools for the cardiac osciallator (see
%   Example~\ref{example:running}).}
% \label{tab:compCardiac}
% \begin{tabular}{|c|c|c|c|}
% \hline
% Tool & Time horizon & Safety & Runtime \\
% \hline
% C2E2 & 20s & Safe & 0.1s\\
% Flow* & 13s & Safe & 20.1s\\
% dReach & N/A & N/A & N/A\\
% CORA & N/A & N/A & N/A\\
% \hline
% \end{tabular}
% \end{center}
% \end{table}

% A comparison for the cardiac oscillator presented earlier as
% Example~\ref{example:running} is shown in Table~\ref{tab:compCardiac}. While
% C2E2 simulates a time horizon of $20$ seconds in very short time,
% Flow* terminates with the error ``The remainder estimation is not large enough"
% at around $13$s which is caused by the big aggregation error for the flowpipes
% over the transition time. dReach and Cora ... \juergen{TODO} The main difference
% of dReach is that it provides only one possible trace that satisfies the end
% condition, not all possible.

%% file: conclusion.tex
%!TEX root = main.tex
\section{Conclusions and Future Work}
\label{sec:conclusion}

%In this paper, we demonstrated that the new C2E2 
%with support for discrepancy computation for systems 
%provides a viable way to add arbitrary external
%Overall, our results confirm that the proposed verification approach with support for discrepancy computation for systems with fixed inputs can indeed be used for simulation and verification of complex non-linear circuit
%models, and in fact opens up new avenues in challenging research areas like metastability analysis.
%

In this paper we introduced a novel approach suitable for
verifying highly sensitive non-linear ODEs
with arbitrary external inputs.
%The new approach avoids the blow-up of the 
%reachtubes in discrepancy-based reachability analysis of systems with inputs considered
%as additional state variables.
%This is mainly achieved by avoiding the blow-up of the 
%reachtubes in discrepancy-based reachability analysis of systems with inputs considered
%as additional state variables.
Its modeling power was shown by successfully implementing several CMOS
circuit models like inverter, {\NOR} gate and memory elements, based on both a
hybrid and a uniform non-linear transistor model
with highly sensitive ODEs and hundreds of nonlinear terms. 
Moreover, we also succeeded to verify the metastable behavior of a memory element, which
demonstrates the ability of our approach to handle highly sensitive ODEs.

The results of this paper  suggest several interesting directions for future research. 
First, for addressing the relatively large simulation time for complex
models, it would be worthwhile to investigate state-of-the-art robust ODE-solvers
for stiff ODEs. Another  direction would be to generalize the core verification
algorithm in order to handle infinite sets of input signals. Finally, we
envision promising applications  in the area of advanced digital circuit
analysis, where C2E2 could be used for  
verifying metastable behavior of circuits like Schmitt-Triggers~\citep{SMN16}.

%% file: main.bbl
\begin{thebibliography}{22}
\providecommand{\natexlab}[1]{#1}
\providecommand{\url}[1]{\texttt{#1}}
\providecommand{\urlprefix}{URL }
\expandafter\ifx\csname urlstyle\endcsname\relax
  \providecommand{\doi}[1]{doi:\discretionary{}{}{}#1}\else
  \providecommand{\doi}{doi:\discretionary{}{}{}\begingroup
  \urlstyle{rm}\Url}\fi

\bibitem[{Althoff(2013)}]{Althoff2013}
Althoff, M. (2013).
\newblock Reachability analysis of nonlinear systems using conservative
  polynomialization and non-convex sets.
\newblock In \emph{Proc. of HSCC '13: the 16th International Conference on
  Hybrid Systems: Computation and Control}, 173--182. ACM.
\newblock \doi{10.1145/2461328.2461358}.

\bibitem[{Althoff and Grebenyuk(2016)}]{Althoff2016a}
Althoff, M. and Grebenyuk, D. (2016).
\newblock Implementation of interval arithmetic in {CORA} 2016.
\newblock In \emph{ARCH Workshop}, 91--105.

\bibitem[{Arora(1993)}]{arora1993}
Arora, N. (1993).
\newblock \emph{MOSFET models for VLSI circuit simulation; theory and
  practice}.
\newblock Computational microelectronics. Springer.

\bibitem[{Chen et~al.(2013)Chen, {\'A}brah{\'a}m, and
  Sankaranarayanan}]{chen2013flow}
Chen, X., {\'A}brah{\'a}m, E., and Sankaranarayanan, S. (2013).
\newblock Flow*: An analyzer for non-linear hybrid systems.
\newblock In \emph{CAV}, 258--263.

\bibitem[{Dang et~al.(2004)Dang, Donz{\'e}, and Maler}]{DDM04:FMCAD}
Dang, T., Donz{\'e}, A., and Maler, O. (2004).
\newblock Verification of analog and mixed-signal circuits using hybrid system
  techniques.
\newblock In \emph{FMCAD}, 21--36.
\newblock \doi{10.1007/978-3-540-30494-4_3}.

\bibitem[{Dang et~al.(2009)Dang, {Le~{G}uernic}, and Maler}]{DangGM09}
Dang, T., {Le~{G}uernic}, C., and Maler, O. (2009).
\newblock Computing reachable states for nonlinear biological models.
\newblock In \emph{CMSB}, volume 5688 of \emph{LNCS}, 126--141. Springer.
\newblock \doi{10.1007/978-3-642-03845-7_9}.

\bibitem[{Donz{\'e}(2010)}]{donze2010breach}
Donz{\'e}, A. (2010).
\newblock Breach, a toolbox for verification and parameter synthesis of hybrid
  systems.
\newblock In \emph{CAV}, 167--170.

\bibitem[{Duggirala et~al.(2013)Duggirala, Mitra, and
  Viswanathan}]{duggirala2013verification}
Duggirala, P.S., Mitra, S., and Viswanathan, M. (2013).
\newblock Verification of annotated models from executions.
\newblock In \emph{EMSOFT}, 1--10.

\bibitem[{Fan and Mitra(2015)}]{chuchuATVA}
Fan, C. and Mitra, S. (2015).
\newblock Bounded verification with on-the-fly discrepancy computation.
\newblock In \emph{ATVA}, 446--463.

\bibitem[{Fan et~al.(2016)Fan, Qi, Mitra, Viswanathan, and Duggirala}]{CAVtool}
Fan, C., Qi, B., Mitra, S., Viswanathan, M., and Duggirala, P.S. (2016).
\newblock Automatic reachability analysis for nonlinear hybrid models with
  {C2E2}.
\newblock In \emph{CAV}, 531--538. Springer.

\bibitem[{Fr{\"{a}}nzle et~al.(2007)Fr{\"{a}}nzle, Herde, Teige, Ratschan, and
  Schubert}]{FranzleHTRS07}
Fr{\"{a}}nzle, M., Herde, C., Teige, T., Ratschan, S., and Schubert, T. (2007).
\newblock Efficient solving of large non-linear arithmetic constraint systems
  with complex boolean structure.
\newblock \emph{{JSAT}}, 1(3-4), 209--236.

\bibitem[{Frehse(2008)}]{Fre08}
Frehse, G. (2008).
\newblock Phaver: algorithmic verification of hybrid systems past hytech.
\newblock \emph{International Journal on Software Tools for Technology
  Transfer}, 10(3), 263--279.
\newblock \doi{10.1007/s10009-007-0062-x}.

\bibitem[{Frehse et~al.(2011)Frehse, {Le~{G}uernic}, Donz{\'e}, Cotton, Ray,
  Lebeltel, Ripado, Girard, Dang, and Maler}]{helicopter}
Frehse, G., {Le~{G}uernic}, C., Donz{\'e}, A., Cotton, S., Ray, R., Lebeltel,
  O., Ripado, R., Girard, A., Dang, T., and Maler, O. (2011).
\newblock Space{E}x: Scalable verification of hybrid systems.
\newblock In \emph{CAV}, 379--395.

\bibitem[{F\"{u}gger et~al.(2015)F\"{u}gger, Najvirt, Nowak, and
  Schmid}]{FNNS15:DATE}
F\"{u}gger, M., Najvirt, R., Nowak, T., and Schmid, U. (2015).
\newblock Towards binary circuit models that faithfully capture physical
  solvability.
\newblock In \emph{DATE}, 1455--1460.

\bibitem[{Gupta et~al.(2004)Gupta, Krogh, and Rutenbar}]{GKR04:ICCAD}
Gupta, S., Krogh, B.H., and Rutenbar, R.A. (2004).
\newblock Towards formal verification of analog designs.
\newblock In \emph{ICCAD}, 210--217.
\newblock \doi{10.1109/ICCAD.2004.1382573}.
\newblock \urlprefix\url{http://dx.doi.org/10.1109/ICCAD.2004.1382573}.

\bibitem[{Henzinger et~al.(1997)Henzinger, Ho, and Wong-Toi}]{henzho:Hytech}
Henzinger, T.A., Ho, P.H., and Wong-Toi, H. (1997).
\newblock Hytech: A model checker for hybrid systems.
\newblock In \emph{CAV}, 460--463. Springer.

\bibitem[{Kong et~al.(2015)Kong, Gao, Chen, and Clarke}]{kong2015dreach}
Kong, S., Gao, S., Chen, W., and Clarke, E. (2015).
\newblock d{R}each: $\delta$-reachability analysis for hybrid systems.
\newblock In \emph{TACAS}, 200--205.

\bibitem[{Lata and Jamadagni(2010)}]{Lata2010}
Lata, K. and Jamadagni, H.S. (2010).
\newblock Formal verification of tunnel diode oscillator with temperature
  variations.
\newblock In \emph{ASPDAC}, 217--222. IEEE Press.

\bibitem[{Maier(2017)}]{Mai17}
Maier, J. (2017).
\newblock Modeling the {CMOS} inverter using hybrid systems.
\newblock Technical Report TUW-259633, E182-TI; TU Wien.
\newblock \urlprefix\url{http://publik.tuwien.ac.at/files/publik_259633.pdf}.

\bibitem[{Marino(1981)}]{Mar81}
Marino, L.R. (1981).
\newblock General theory of metastable operation.
\newblock \emph{IEEE ToC}, 30(2), 107--115.

\bibitem[{Steininger et~al.(2016)Steininger, Maier, and Najvirt}]{SMN16}
Steininger, A., Maier, J., and Najvirt, R. (2016).
\newblock The metastable behavior of a {S}chmitt-{T}rigger.
\newblock In \emph{ASYNC}, 57--64.
\newblock \doi{10.1109/ASYNC.2016.19}.

\bibitem[{Yan and Greenstreet(2008)}]{coho2008}
Yan, C. and Greenstreet, M.R. (2008).
\newblock Verifying an arbiter circuit.
\newblock In \emph{FMCAD}, 7:1--7:9. IEEE Press, Piscataway, NJ, USA.
\newblock \urlprefix\url{http://dl.acm.org/citation.cfm?id=1517424.1517431}.

\end{thebibliography}
